\documentclass[12pt,a4paper]{scrartcl}

\usepackage{amsmath,amssymb,amsthm}
\usepackage{graphicx}
\usepackage{ulem}

\usepackage{times}
\usepackage{bm}
\usepackage{natbib}
\usepackage{xcolor}
\usepackage[plain,noend]{algorithm2e}

\newtheorem{theorem}{Theorem}
\newtheorem{corollary}[theorem]{Corollary}

\newtheorem{proposition}[theorem]{Proposition}
\newtheorem{assumption}[theorem]{Assumption}

\theoremstyle{definition}
\newtheorem{example}[theorem]{Example}
\newtheorem{definition}[theorem]{Definition}
\newtheorem{remark}[theorem]{Remark}

\def\N{\mathbb{N}}
\def\R{\mathbb{R}}
\def\supp{\operatorname{supp}}
\def\int{\operatorname{int}}
\def\W{\operatorname{W}}

\def\Rdists{\mathcal{P}}
\def\Sdists{\mathcal{S}}

\def\SAS{\operatorname{SAS}}

\allowdisplaybreaks

\begin{document}
\title{Centre-free kurtosis orderings for asymmetric distributions}

\author{Andreas Eberl\footnote{andreas.eberl@kit.edu},
Bernhard Klar\footnote{ bernhard.klar@kit.edu} \\
\small{Institute of Stochastics,} \\
\small{Karlsruhe Institute of Technology (KIT), Germany.}
}

\date{\today }
\maketitle

\begin{abstract}
The concept of kurtosis is used to describe and compare theoretical and empirical distributions in a multitude of applications. In this connection, it is commonly applied to asymmetric distributions. However, there is no rigorous mathematical foundation establishing what is meant by kurtosis of an asymmetric distribution and what is required to measure it properly. All corresponding proposals in the literature centre the comparison with respect to kurtosis around some measure of central location. Since this either disregards critical amounts of information or is too restrictive, we instead revisit a canonical approach that has barely received any attention in the literature. It reveals the non-transitivity of kurtosis orderings due to an intrinsic entanglement of kurtosis and skewness as the underlying problem. This is circumvented by restricting attention to sets of distributions with equal skewness, on which the proposed kurtosis ordering is shown to be transitive. Moreover, we introduce a functional that preserves this order for arbitrary asymmetric distributions.
As application, we examine the families of Weibull and sinh-arsinh distributions and show that the latter family exhibits a skewness-invariant kurtosis behaviour.
%Another transitivity set is given in the form of a specific distribution family with skewness-invariant kurtosis behaviour.
\end{abstract}

{\bfseries Keywords:} Asymmetric distribution; Higher-order convexity; Kurtosis; Skewness; Stochastic order; Sinh-arsinh distribution.

\section{Introduction}
\label{sec:intro}

There has been much discussion in the literature concerning the question of what kurtosis describes exactly. In particular, a number of articles have been published both advocating its interpretation as 'peakedness' of a distributions and opposing it. See \citet[pp.\ 72--79]{crack} and \citet{rip_kurtPeak} for examples of either position and \citet{kurtHistRevw} for a more neutral historical review. \citet[p.\ 116]{MacG_kurt_revw} provide a critical review of the literature concerning kurtosis and, based on that, aptly describe an increase in kurtosis as 'the location- and scale-free movement of probability mass from the shoulders of a distribution into its center and tails'. This heuristic, as is usually the case for kurtosis, is applied solely to unimodal symmetric distributions.

As for other distributional characteristics, the concept of kurtosis is usually rooted in a stochastic order. Among the first authors to introduce this order-based approach for location and dispersion were \cite{bl2,bl3}, it was later generalized by \citet{oja}, among others. In particular, they required any measure $\nu$ of a specific characteristic of a distribution to preserve a corresponding order $\preccurlyeq$, i.e.\ that $F \preccurlyeq G$ implies $\nu(F) \leq \nu(G)$ for all sufficiently regular distribution functions $F, G$. The necessity of underpinning measures in this way is, e.g., demonstrated in \citet{ek-patil}. The most popular choice in the literature for this fundamental stochastic order in the case of kurtosis was  introduced by \citet{zwet} and is denoted by $\leq_s$. Two distributions $F, G$ are said to be ordered with respect to kurtosis in the sense of $F \leq_s G$, if the function $x \mapsto R_{FG}(x) = G^{-1}(F(x))$ is convex for $x \geq F^{-1}(1/2)$. Again, this fundamental order is only meaningful if $F$ and $G$ are symmetric.

Although not well-founded, the application of kurtosis and, more specifically, measures of kurtosis to asymmetric distribution is commonplace: when proposing new families of continuous distributions or methods for generating such families, shape parameters are often related to skewness and kurtosis. Examples are \citet{goerg:2011}, \citet{Alzaatreh:2013} and \citet{Fischer:2013}.
Only occasionally, authors are more reluctant: \cite{Jones:2009} term the second shape parameter of their sinh-arsinh distribution as kurtosis only in the symmetric case, otherwise they speak about tailweight, well aware of the underlying subtleties.

When modeling stock market volatility, \citet{Gabaix:2006} consider distributions with large values of moment based skewness and kurtosis and conclude ``The use of [moment] kurtosis should be banished from use with fat-tailed distributions.''
\citet{Asmussen:2022} studies higher order cumulants for a selection of financial models from the literature. His motivation ``comes from numerous statements in the financial literature in the spirit that S [skewness] accounts for asymmetry and K [kurtosis] for a sharper mode and heavier tails than for the Black-Scholes model.''

In particular in applied work, the notion of kurtosis is routinely used for skewed distributions, and sample skewness and kurtosis are frequently documented in the literature. Examples are \citet{Bai:2005, Kim:2004} in the context of modeling financial returns; \citet{Szczygielski:2020, Lopez-Martin:2022} for modeling cryptocurrencies; \citet{Martins:1965, Cooper:2020} for environmental data; \citet{Eling:2012, Sherrick:2004} in the context of insurance risk.

All major approaches in the literature to define a fundamental kurtosis order for asymmetric distributions have the same critical drawback. Namely, they artificially centre the comparison of two distributions with respect to kurtosis around some measure of location, usually the median. Examples include the anti-skewness order $\leq_a$ by \citet{MacG_kurt_skew} and the order $\leq_S$ by \citet{MacG_kurt_spread}, which is based on the so-called spread function. The order $\leq_a$ basically imposes the same requirement as $\leq_s$ since $F \leq_a G$ is equivalent to $R_{FG}$ being concave up to the median of $F$ and convex from there onward. While this switch necessarily takes place at the median in a symmetric setting, it is very limiting and not expedient to require this in the general case. A more flexible generalization of the concave-convex order $\leq_s$ is proposed in Section \ref{sec:leqgs}.

For the second order $\leq_S$, the spread function of a distribution function $F$ is defined by
\begin{equation*}
	S_F(\alpha) = F^{-1}(\tfrac{1}{2}+\alpha) - F^{-1}(\tfrac{1}{2}-\alpha), \quad \alpha \in [0, \tfrac{1}{2}),
\end{equation*}
which can be interpreted as one half of a symmetrized quantile function of $F$. Heuristically, the distribution is again artificially centred around the median by folding it around the median and averaging out the two overlaying halves of the distributions. If the resulting half of a distribution is then mirrored at the median, a symmetric distribution is obtained, which can be ordered with respect to kurtosis using $\leq_s$. This methodology is equivalent to defining the symmetrized kurtosis order $\leq_S$ by
\begin{equation*}
%\label{eqn:defLeqS}
	F \leq_S G \quad \Leftrightarrow \quad S_G \circ S_F^{-1} \text{ is convex}.
\end{equation*}
This definition of a kurtosis order is fairly easy to use and theoretically applicable to all univariate distributions. It does, however, have significant downsides, especially if it is intended to be used as a foundational order that establishes what is meant by the notion of kurtosis. This was in part noted by \citet[p.\ 29]{MacG_kurt_spread} themselves. First, a significant amount of information is lost in just combining the two 'sides' (with respect to the median) of the distribution. The order $\leq_S$ theoretically allows arbitrarily large deviations from the desired concavity or convexity on one side, if they are compensated by the other side. This kind of behaviour is not desirable for a basic order. In a financial context, for instance, negative and positive values of the distribution, i.e.\ losses and gains, have to be interpreted differently, and relevant information about the shape of the distribution is lost by forcing symmetry. The second downside becomes apparent if we consider skewed distributions on the positive half line. In this case, the support ends close to the median on one side and is infinite on the other side, and the symmetrized version is not representative of the original distribution.

Further proposals of kurtosis orders for asymmetric distributions were discussed by \citet{oja}, \citet{MacG_kurt_revw}, \citet{ag_kurt_orders} and \citet{fiori_kurt_order}, but they exhibit similar drawbacks to $\leq_a$ and $\leq_S$.

\citet[p.\ 168]{oja} also briefly mentions the kurtosis order $\leq_3$, where $F \leq_3 G$ is said to hold if $R_{FG}$ is convex of order three. If both $F$ and $G$ are three times differentiable, this is equivalent to $R_{FG}''' \geq 0$. This definition naturally arises from basic orders of location, dispersion and skewness that are based upon the function $\Delta_{FG}(x) = R_{FG}(x) - x$. These orders, denoted $\leq_0$, $\leq_1$ and $\leq_2$ by \citet{oja}, hold if $\Delta_{FG}$ is non-negative, increasing and convex in the usual sense, respectively. For continuous distributions, $\leq_0$ coincides with the usual stochastic order. Under appropriate differentiability assumptions, the definition of all three orders can be unified by stating that $F \leq_k G~(k = 0, 1, 2)$ holds if $\Delta_{FG}^{(k)} \geq 0$.
Since the concepts of location, dispersion, skewness and kurtosis are hierarchically connected,
as can be seen from the classical measures, the first, second, third and fourth standardized moment, $\leq_3$ seems to be the canonical basic kurtosis order. In particular, $\leq_3$ is naturally applicable to asymmetric distributions.
In spite of these observations, the order $\leq_3$ is otherwise not considered in the literature except by \citet{hosking89}, who shows that his kurtosis measure based on L-moments preserves $\le_3 $ for symmetric distributions. This disregard may partly be due to \citet{oja} criticizing the order for not being transitive.

If the order $\leq_3$ was lacking transitivity on symmetric distributions, this would indeed be a serious downside compared to $\leq_s$. However, in Section \ref{sec:leq3}, it is proved that $\leq_3$ is transitive on this set. For the set of all distributions, we argue that transitivity cannot be expected, since skewness or asymmetry interferes with the quantification of kurtosis.
This intrinsic entanglement was already mentioned by \citet{MacG_kurt_skew} and \citet{MacG_kurt_spread}, and proposals for skewness-invariant kurtosis measures were made by \citet{kurt_ohne_skew_mom} and \citet{jones_quint}. In Sections \ref{sec:leq3} and \ref{sec:leqgs}, this entanglement is shown to be related to the transitivity of kurtosis orders.

%Dann die vorhandenen Ansaetze in der Literatur und warum sie unzureichend sind, anschliessend Outline des Papers (aber: Biometrika papers do not contain a ‘contents of the paper’ paragraph).

\section{The kurtosis order $\boldsymbol{\leq_3}$ and its transitivity properties}
\label{sec:leq3}

\subsection{Basics}

We begin by defining convex functions of order $k \in \N_0$ and the induced stochastic orders.

\begin{definition}
\label{def:kCx}
Let $I \subseteq \R$ be an open interval and let $\varphi: I \to \R$ be a function. For $k \in \N$ and $x_0, \ldots, x_k \in I$ with $x_0 < \ldots < x_k$, the zeroth and $k$-th {\em divided difference}, respectively, of $\varphi$ at $x_0, \ldots, x_k$ is defined by
\begin{align*}
	[x_0|\varphi] &= \varphi(x_0),\\
	[x_0, \ldots, x_k|\varphi] &= \frac{[x_1, \ldots, x_k|\varphi] - [x_0, \ldots, x_{k-1}|\varphi]}{x_k - x_0}.
\end{align*}
$\varphi$ is said to be {\em convex of order $k$} or {\em $k$-convex} on $I$, if
\begin{equation}
\label{eqn:genCx}
	[x_0, \ldots, x_k|\varphi] \geq 0
\end{equation}
holds for all $x_0, \ldots, x_k \in I$ with $x_0 < \ldots < x_k$. Moreover, $\varphi$ is said to be {\em strictly convex of order $k$} on $I$, if inequality (\ref{eqn:genCx}) is strict.
\end{definition}

The $k$-convexity of functions can also be defined via the non-negativity of determinants of $(k+1)\times(k+1)$-matrices \citep[see][p.\ 155]{oja}. It is easy to see that both approaches are equivalent.
Throughout this work, we assume the following.

\begin{assumption}
All (univariate) distribution functions have interval support and are three times differentiable. The interior of the support of a distribution function $F$ is denoted by $D_F$ and $f=F'$ is assumed to be strictly positive on $D_F$. The set of all such distribution functions is denoted by $\Rdists$.
\end{assumption}

\citet[p.\ 156]{oja} defined a family of stochastic orders in the following way.

\begin{definition}
\label{def:leqk}
Let $k \in \N_0$ and $F, G \in \Rdists$. Then, $F \leq_k G$ is said to hold, if the function
\begin{equation*}
    \Delta_{FG}: D_F \to \R, \quad x \mapsto R_{FG}(x) - x = G^{-1}(F(x)) - x
\end{equation*}
is convex of order $k$.
\end{definition}

Here, $G^{-1}$ denotes both the inverse function and the quantile function of $G$, which coincide given our regularity conditions. Note that $\leq_0$ coincides with the usual stochastic order $\leq_{st}$, the most basic location order. Similarly, $\leq_1$ coincides with the basic dispersion order $\leq_{disp}$ and $\leq_2$ coincides with the basic skewness order $\leq_c$ by \citet{zwet}. For more details, see \citet{oja}. Since the $k$-convexity of a $k$ times differentiable function $\varphi$ is equivalent to $\varphi^{(k)} \geq 0$, one obtains the following corollary \citep[see, e.g.,][]{oja}.

\begin{corollary}
Let $k \in \N_0$ and $F, G \in \Rdists$. Then, $F \leq_k G$ is equivalent to $\Delta_{FG}^{(k)} \geq 0$. If $k \geq 2$, $F \leq_k G$ is also equivalent to $R_{FG}^{(k)} \geq 0$.
\end{corollary}

\subsection{Lack of transitivity and its implications}

We now focus our attention on the order $\leq_3$ as a canonical choice for a basic kurtosis order. In a rare mention in the literature, \citet[p.\ 168]{oja} states without proof that $\leq_3$ is not transitive. This is confirmed by the following example.

\begin{example}
\label{exm:CounterTransLeq3}
%     \begin{figure}
% 		\centering{
% 			\includegraphics[scale=0.5]{CEX_leq3_trans}
% 			\caption{\label{fig:CEX_leq3_trans}Graphs of the three cdf's (left panel) and the three RIDF's from Example \ref{exm:CounterTransLeq3}.}
% 		}
% 	\end{figure}
	Define by
	\begin{align*}
		F: [0, 1] \to [0, 1],&\quad t \mapsto t^3,\\
		G: [0, 1] \to [0, 1],&\quad t \mapsto t,\\
		H: [0, 1] \to [0, 1],&\quad t \mapsto 1 - (1-t)^{1/3}
	\end{align*}
	three infinitely often differentiable distribution functions on the unit interval. Note that $D_F = D_G = D_H = (0, 1)$. Since both $R_{FG}$ and $R_{GH}$ are shifted versions of the third monomial with restricted domains, both $F \leq_3 G$ and $G \leq_3 H$ hold. Straightforward calculations yield
	\begin{align*}
		R_{FH}'''(t) &= 18(28t^6-20t^3+1),
	\end{align*}
	which implies
	\begin{align*}
		R_{FH}'''(t) < 0 \text{ for } t \in \left( \left(\tfrac{5-3\surd{2}}{14}\right)^{1/3}, \left(\tfrac{5+3\surd{2}}{14}\right)^{1/3} \right) \approx (0.378, 0.871) \subseteq [0, 1],
	\end{align*}
	and thereby contradicts $F \leq_3 H$.
\end{example}

The orders $\leq_0$, $\leq_1$ and $\leq_2$ can all be equivalently characterized by families of measures of location, dispersion and skewness, respectively. Because all such measures are mappings from a set of probability distributions to the real numbers, their values are compared using the transitive relation $\leq$. Since this is not compatible with the non-transitivity of the kurtosis order $\leq_3$, we obtain the following negative result.

\begin{corollary}
\label{thm:noKurtMeasFam}
	There does not exist a family $\{\kappa_{\iota}: \Rdists \to \R \,|\, \iota \in I\}$ of mappings such that
	\begin{align*}
		\kappa_{\iota}(F) \leq \kappa_{\iota}(G) \quad \forall \iota\in I
	\end{align*}
	is equivalent to $F\leq_3 G$.
\end{corollary}

Note that stochastic orders are usually not strongly connected, which means that $F \not\leq_3 H$ does not imply $H \leq_3 F$. However, for the distributions in Example \ref{exm:CounterTransLeq3}, it can be shown that $H \leq_3 F$ does indeed hold. This is a more disturbing result than the mere non-transitivity of $\leq_3$: if one additionally requires that $\kappa$ preserves the strict version of $\leq_3$, i.e.\ that $F <_3 G$ implies $\kappa(F) < \kappa(G)$, it can be shown that there exists no mapping $\kappa: \Rdists \to \R$ that preserves the order $\leq_3$.

%Bei der Diskussion der Transitivität:
\begin{remark} \label{u-test}
It should be emphasized that missing transitivity can also be found in more familiar areas. The best known example is probably the location order defined by $X \leq_p Y$ if the relative effect $p=P(X<Y)+P(X=Y)/2$ is greater than or equal to $1/2$. It is well known that this order is not transitive, as exemplified by non-transitive dice \citep{Gardner:1970}. Still, the empirical counterpart to $p$ is the key quantity of important nonparametric tests like the Wilcoxon-Mann-Whitney, Fligner-Policello and Brunner-Munzel test \citep{Divine:2018}.

An example for a non-transitive dispersion ordering is the dangerousness order: given random variables $X,Y$ on the positive half line, $X$ is said to be less dangerous than $Y$ if there is some $c\geq 0$ with $F\leq G$ on $[0,c)$, $F\geq G$ on $[c,\infty)$ and $E(X)\leq E(Y)$. Here, the situation somewhat differs from the foregoing  example, since the dangerousness order has a transitive closure, the convex order \citep{Mueller:1996}.
\end{remark}

The observation preceding Remark \ref{u-test} suggests that kurtosis measures in the classical order-based sense, used by \citet{oja} among others, do not exist. In the literature, $\leq_s$ is usually chosen as the kurtosis order to be preserved by a kurtosis measure. Because of the limitations of $\leq_s$, this can only be used to validate kurtosis measures for symmetric distributions, which is unsatisfactory for the reasons mentioned in Section \ref{sec:intro}.
The  question how to use the much more general applicability of the order $\leq_3$ in spite of its non-transitivity can be answered in two ways.

The first possibility is to move away from the classical idea of measures of kurtosis and instead consider functionals that quantify the difference in kurtosis between two given distributions. For example, consider the quantile-based mapping
\begin{align*}
	\kappa_Q^{\alpha,\eta}: \Rdists \to \R, \quad F \mapsto \frac{F^{-1}(1-\alpha)-3F^{-1}(1-\eta)+3F^{-1}(\eta)-F^{-1}(\alpha)}{F^{-1}(1-\eta)-F^{-1}(\eta)}
\end{align*}
for $0 < \alpha < \eta < 1/2$, which is listed as a kurtosis measure by \citet{ruppert}, \citet{MacG_kurt_revw} and \citet{jones_quint} among others, since it preserves the order $\leq_s$. Similarly constructed quantile-based mappings using lower-order differences are measures of location, dispersion and skewness and can even be used to characterize the orders $\leq_0$, $\leq_1$ and $\leq_2$ in the sense of Corollary \ref{thm:noKurtMeasFam}. By customizing the evaluation points to a second distribution, one arrives at the functional
\begin{align*}
%\label{eqn:kappaQFDef}
	\kappa_{QF}^\alpha(F,G) =
	\frac{G^{-1}(1-\alpha) - 3G^{-1}\left(\eta_F(1-\alpha)\right)
	+ 3G^{-1}\left(\eta_F(\alpha)\right) - G^{-1}(\alpha)}{G^{-1}(1-\alpha) - G^{-1}(\alpha)}, \; \; 0<\alpha<\tfrac{1}{2},
\end{align*}
where
\begin{align*}
	\eta_F(q) = F\left(\tfrac{2}{3}F^{-1}(q)+\tfrac{1}{3}F^{-1}(1-q)\right).
\end{align*}
This functional preserves the kurtosis order $\leq_3$ even for asymmetric distributions, as the following result shows.

\begin{proposition}
Let $F, G \in \Rdists$. Then $F \leq_3 G$ implies $\kappa_{QF}^\alpha(F, G) \geq 0$.
\begin{proof}
According to Definitions \ref{def:kCx} and \ref{def:leqk}, $F \leq_3 G$ is equivalent to
\begin{align}
\label{eqn:divDifKurt}
	\frac{ \frac{G^{-1}(p_3)-G^{-1}(p_2)}{F^{-1}(p_3)-F^{-1}(p_2)}
		-\frac{G^{-1}(p_2)-G^{-1}(p_1)}{F^{-1}(p_2)-F^{-1}(p_1)} }{F^{-1}(p_3)-F^{-1}(p_1)}
	-\frac{ \frac{G^{-1}(p_2)-G^{-1}(p_1)}{F^{-1}(p_2)-F^{-1}(p_1)}
		-\frac{G^{-1}(p_1)-G^{-1}(p_0)}{F^{-1}(p_1)-F^{-1}(p_0)} }{F^{-1}(p_2)-F^{-1}(p_0)}
	&\geq 0
\end{align}
for all $0 < p_0 < p_1 < p_2 < p_3 < 1$. By choosing $p_0 = \alpha$, $p_1 = \eta_F(\alpha)$, $p_2 = \eta_F(1-\alpha)$ and $p_3 = 1 - \alpha$, (\ref{eqn:divDifKurt}) boils down to
\begin{align*}
	G^{-1}(1-\alpha)-3G^{-1}(\eta_F(1-\alpha))+3G^{-1}(\eta_F(\alpha))-G^{-1}(\alpha)\geq 0.
\end{align*}
The subsequent division by the $\alpha$-interquantile range of $G$ is done to obtain a scale-invariant functional.
\end{proof}
\end{proposition}

%Nach der Einführung des Funktionals:
\begin{remark}
Again, the situation is similar for the Wilcoxon-Mann-Whitney location order $\leq_p$ in Remark \ref{u-test}. The usual unbiased estimator for the relative effect is a U-statistic involving both samples, and there cannot exist a measure depending on one sample like the mean or median, which is consistent with $\leq_p$ in general.
\end{remark}

\subsection{Transitivity sets}
\label{sec:leq3TS}

The second possibility is to restrict the comparison of kurtosis to suitable subsets of distributions, e.g.\ the subset of symmetric distributions. In the following, we analyze the transitivity sets of the order $\leq_3$.
As a starting point, all pairs of distributions that are ordered with respect to $\leq_3$ are divided into two mutually exclusive categories. For that, let $F, G \in \Rdists$ satisfy $F \leq_3 G$, implying that the function $R_{FG}''$ is increasing. Now, $F$ and $G$ are either skewness-comparable with respect to $\leq_2$, i.e., $F \leq_2 G$ or $G \leq_2 F$ holds, or they are not. In the latter case, $R_{FG}$ has an inflection point at a $t_{FG} \in D_F = \int(\supp(F))$ with $R_{FG}''(t) \leq 0$ for $t \leq t_{FG}$ and $R_{FG}''(t) \geq 0$ for $t \geq t_{FG}$. More specifically, there exist values $t_\ell, t_u \in D_F$ with $t_\ell < t_{FG} < t_u$ such that $R_{FG}''(t_\ell) < 0$ and $R_{FG}''(t_u) > 0$.
The inflection point at $t_{FG}$ is, in general, not unique since $R_{FG}$ can be linear on a given non-degenerate interval. However, any inflection point of $R_{FG}$ can be uniquely identified by the value $p_{FG} = F(t_{FG}) \in (0, 1)$.
Furthermore, note that $F \leq_2 G$ and $G \leq_2 F$ can be viewed as limiting cases with $t_{FG} = \inf D_F$ or $t_{FG} = \sup D_F$, yielding $p_{FG} = 0$ or $p_{FG} = 1$, respectively. So in order to obtain the most general setting, we allow $t_{FG} \in \overline{D_F} = \supp(F)$.

\begin{definition}
\label{def:inflVal}
	Let $F$ and $G$ be two cdf's satisfying $F \leq_3 G$. A value $p_{FG} \in [0, 1]$ is said to be an {\em inflection value of $F$ and $G$}, if $R_{FG}''(t) \leq 0$ for all $t \leq F^{-1}(p_{FG})$ and $R_{FG}''(t) \geq 0$ for all $t \geq F^{-1}(p_{FG})$. The set of all inflection values of $F$ and $G$ is denoted by $\Pi_{FG}$.
\end{definition}

%With this definition, a pair $F, G$ of cdf's can fall into case (i) in Remark \ref{rmk:TransLeq3Tp} and still have an inflection value $p_{FG} \in (0, 1)$. This holds, if and only if there exists a $t_0 \in D_F$ such that $R_{FG}$ is linear on $(\inf D_F, t_0)$ or $(t_0, \sup D_F)$.\par
As stated before, any pair $F, G$ satisfying $F \leq_3 G$ has at least one inflection value. Requiring $R_{FG}'''(t) > 0$ for all $t \in D_F$ is sufficient for the inflection value $p_{FG}$ to be unique.
With this in mind, we analyze more closely why $\leq_3$ is not transitive. Let $F, G$ and $H$ satisfy $F \leq_3 G$ and $G \leq_3 H$. Then,
\begin{align}
\nonumber
    R_{FH}(t) =& \; H^{-1}(F(t)) = H^{-1}(G(G^{-1}(F(t)))) = R_{GH}(R_{FG}(t)),\\
% \nonumber
% 	R_{FH}'(t) =& \; R_{GH}'(R_{FG}(t)) \cdot R_{FG}'(t),\\
\label{eqn:2ndDerivativeDeltaFH}
	R_{FH}''(t) =& \; R_{GH}''(R_{FG}(t)) \cdot (R_{FG}'(t))^2 + R_{GH}'(R_{FG}(t)) \cdot R_{FG}''(t),\\
\nonumber
	R_{FH}'''(t) =& \; R_{GH}'''(R_{FG}(t)) \cdot (R_{FG}'(t))^3 + R_{GH}'(R_{FG}(t)) \cdot R_{FG}'''(t)\\
\label{eqn:3rdDerivativeDeltaFH}
	& \; + 3 R_{GH}''(R_{FG}(t)) \cdot R_{FG}'(t) \cdot  R_{FG}''(t)
	%	& \; + R_{GH}''(R_{FG}(t)) \cdot R_{FG}''(t) \cdot \left( 1 + 2 R_{FG}'(t) \right).
\end{align}
holds for all $t \in D_F$. Note that $R_{FG}$ and $R_{GH}$ are increasing as a composition of two increasing functions. Hence, the first two summands on the right side of equation (\ref{eqn:3rdDerivativeDeltaFH}) are non-negative and
\begin{equation*}
%\label{eqn:TransSuffCond}
	R_{GH}''(G^{-1}(p)) \cdot R_{FG}''(F^{-1}(p)) \geq 0 \quad \text{for all } p \in (0, 1)
\end{equation*}
is a sufficient condition for $F \leq_3 H$. By assumption, the sets $\Pi_{FG}$ and $\Pi_{GH}$ are both non-empty. If the intersection of these two sets is also non-empty, i.e., if there exists a $p_0 \in [0, 1]$ such that $p_0 \in \Pi_{FG}$ and $p_0 \in \Pi_{GH}$, the signs of $R_{FG}''(F^{-1}(p))$ and $R_{GH}''(G^{-1}(p))$ coincide for all $p \in (0, 1)$ since they are both non-positive for $p < p_0$ and both non-negative for $p > p_0$. Otherwise, if the intersection of $\Pi_{FG}$ and $\Pi_{GH}$ is empty, choose a representative from each set such that their difference is minimal. Assuming without restriction that $p_{FG} < p_{GH}$, where $p_{FG} \in \Pi_{FG}$ and $p_{GH} \in \Pi_{GH}$, it follows that
\begin{equation*}
	R_{GH}''(G^{-1}(p)) \cdot R_{FG}''(F^{-1}(p)) < 0 \quad \text{for all } p \in (p_{FG}, p_{GH}).
\end{equation*}
We summarize our results thus far in the following proposition.

\begin{proposition}
	\label{thm:sameIV_leq3Trans}
	Let $p_0 \in [0, 1]$ and let $\mathcal{F}_0$ be a set of cdf's such that any pair $F, G \in \mathcal{F}_0$ with $F \leq_3 G$ has $p_0$ as an inflection value. Then, the order $\leq_3$ is transitive on $\mathcal{F}_0$.
\end{proposition}

%Overall, (\ref{eqn:TransSuffCond}) is satisfied
%(which is, in turn, a sufficient condition for $F \leq_3 H$)
%if and only if the pairs $F, G$ and $G, H$ have coinciding inflection values. Hence, $\leq_3$ is transitive on sets, where the relative inverse distribution functions of all possible pairs have the same inflection point.
We now study the structure of the sets mentioned in Proposition \ref{thm:sameIV_leq3Trans} or suitable subsets thereof.
First, we assume that $F$ and $G$ with $F \leq_3 G$ have an inflection value $p_{FG} \in (0, 1)$.
The fact that $p_{FG} = F(t_{FG}) \in (0, 1)$ is an inflection value of the pair $F, G$ is equivalent to $R_{FG}''(t_{FG}) = 0$.
%Considering $\Delta_{FG} = G^{-1} \circ F$ as well as
Denoting by $f$ and $g$ the derivatives of $F$ and $G$, respectively, we get
\begin{align}
% 	\label{eqn:1stDerRFG}
% 	R_{FG}'(t) &= \frac{f(t)}{g(R_{FG}(t))},\\
	\label{eqn:2ndDerRFG}
	R_{FG}''(t) &= \frac{f'(t) \cdot (g(R_{FG}(t)))^2 - f^2(t) \cdot g'(R_{FG}(t))}{(g(R_{FG}(t)))^3}
\end{align}
for all $t \in D_F$. Hence, $p_{FG}$ is an inflection value of $F$ and $G$, if and only if
\begin{align}
% \nonumber
% 	& \;& f'(t_{FG}) \cdot (g(R_{FG}(t_{FG})))^2 &=& \; &(f(t_{FG}))^2 \cdot g'(R_{FG}(t_{FG}))\\
% \nonumber
% 	\Longleftrightarrow& \;& \frac{f'(t_{FG})}{(f(t_{FG}))^2} &=& \; &\frac{g'(R_{FG}(t_{FG}))}{(g(R_{FG}(t_{FG})))^2}\\
\label{eqn:EquivCondInflVal}
	%\Longleftrightarrow& \;&
	\frac{f'(F^{-1}(p_{FG}))}{(f(F^{-1}(p_{FG})))^2} &= \frac{g'(G^{-1}(p_{FG}))}{(g(G^{-1}(p_{FG})))^2}.
\end{align}
Thus, any pair that is ordered with respect to $\leq_3$ out of a given set of cdf's has the same inflection value $p_0 \in (0, 1)$, if and only if
$$\gamma_D^{p_0}(F) = \frac{f'(F^{-1}(p_0))}{(f(F^{-1}(p_0)))^2}$$
coincides for all cdf's $F$ in the set. The following result is obtained by combining this observation with Proposition \ref{thm:sameIV_leq3Trans}.

\begin{proposition}
\label{thm:constDensSkew_leq3Trans}
	Let $p_0 \in (0, 1)$ and let $\mathcal{F}_0$ be a set of cdf's such that  $\gamma_D^{p_0}(F)$ coincides for all $F \in \mathcal{F}_0$. Then, all pairs $F, G \in \mathcal{F}_0$ with $F \leq_3 G$ have $p_0$ as an inflection value.
	%and, by Proposition \ref{thm:sameIV_leq3Trans}, the order $\leq_3$ is transitive on $\mathcal{F}_0$.
\end{proposition}

If $p_{FG} \in \{0, 1\}$ is the sole inflection value of $F$ and $G$ with $F \leq_3 G$, (\ref{eqn:2ndDerRFG}) is not valid because the densities $f$ and $g$ are not uniquely defined at the edges of their respective supports. Thus, no easily verifiable sufficient condition for inflection points as in Proposition \ref{thm:constDensSkew_leq3Trans} can be obtained in this case.
In summary, defining the set
\begin{equation*}
	\mathcal{T}_{D, p}^t = \{ F \in \Rdists: \gamma_D^p(F) = t \}
\end{equation*}
for all $p \in (0, 1)$ and all $t \in \R$ gives the following result.

% For $p \in (0, 1)$, the quantities $\gamma_D^p(F) = \frac{f'(F^{-1}(p))}{(f(F^{-1}(p)))^2}$ have already been defined in Section \ref{sec:LDSDensMeas}. They satisfy the crucial condition (S2) for skewness measures for all $p \in (0, 1)$, but $p = \frac{1}{2}$ is the only choice for which $\gamma_D^p$ is a skewness measure (see Corollary \ref{thm:LDSDensMeasP}). By defining the set
% \begin{equation*}
% 	\mathcal{T}_{D, p}^t = \{ F \in \ICdists{3}: \gamma_D^p(F) = t \}
% \end{equation*}
% for all $p \in (0, 1)$ and all $t \in \R$ and defining $\mathcal{T}_D^t = \mathcal{T}_{D, \frac{1}{2}}^t$, we can rephrase Proposition \ref{thm:constDensSkew_leq3Trans} in the following way.

\begin{theorem}
\label{thm:kurtDensTrans}
	For any $t \in \R$ and any $p \in (0, 1)$, the kurtosis order $\leq_3$ is transitive on the set $\mathcal{T}_{D, p}^t$.
\end{theorem}

As mentioned in Section \ref{sec:intro}, a number of authors have identified an intrinsic entanglement between skewness and kurtosis. By considering the mapping $\gamma_D^p$ more closely, this observation is confirmed and refined by Theorem \ref{thm:kurtDensTrans}. Recall that the critical property of a skewness measure $\gamma: \Rdists \to \R$ is that it preserves the skewness order $\leq_2$, i.e.\ that $F \leq_2 G$ implies $\gamma(F) \leq \gamma(G)$ for all $F, G \in \Rdists$. Since $F \leq_2 G$ is equivalent to $R_{FG}'' \geq 0$, changing equations (\ref{eqn:2ndDerRFG}) and (\ref{eqn:EquivCondInflVal}) into inequalities yields that $\gamma_D^p$ preserves $\leq_2$ for all $p \in (0, 1)$ and thus measures skewness. In fact, $\gamma_D^p(F) \leq \gamma_D^p(G)$ for all $p \in (0, 1)$ is equivalent to $F \leq_2 G$, so these measures characterize the order $\leq_2$ in a way that is not possible for $\leq_3$ according to Corollary \ref{thm:noKurtMeasFam}. However, for $p \neq 1/2$, $\gamma_D^p$ measures skewness in an asymmetric or non-central way because the additional requirement $\gamma_D^p(-X) = -\gamma_D^p(X)$ \citep[see, e.g.,][p.\ 393]{groeneveld} is not satisfied.

The fact that $\leq_3$ is transitive, if a suitable skewness measure is constant, suggests that the non-transitivity of $\leq_3$ on the set of all cdf's is because pairs of cdf's with differing degrees of skewness lack comparability with respect to kurtosis.
%\textcolor{red}{Since $\gamma_D^p$ only measures skewness in a symmetric way for $p = 1/2$, it is also notable that, for all $p \in (0, 1)$, $p$ is an inflection value of all pairs $F, G \in \mathcal{T}_{D, p}^t$ with $F \leq_3 G$. Thus, the inflection values for cdf's in $\mathcal{T}_{D, p}^t$ lie in the centre of the unit interval if and only if $p = 1/2$. [später? Formulierung?]}
%
As opposed to location and dispersion, a distribution cannot be standardized with respect to skewness by an arithmetic operation like addition for location and scalar multiplication for dispersion. Thus, in order to obtain a transitive kurtosis order without interference caused by skewness, attention has to be restricted to sets of constant skewness.  Note that, for all $p \in (0, 1)$, the sets $\mathcal{T}_{D, p}^t, t \in \R,$ constitute an partition of the set $\Rdists$ of distributions. For each partition, $p$ is also the inflection value of every kurtosis comparable pair of distributions from the same transitivity set of the partition.
%$\bigcup_{t \in \R} \mathcal{T}_{D, p}^t = \Rdists$ holds because all distributions are assigned a value by the measure $\gamma_D^p$, and $\mathcal{T}_{D, p}^t \cap \mathcal{T}_{D, p}^{s} = \emptyset$ holds for all $s, t \in \R, s \neq t,$ because one distribution cannot be assigned multiple values by that measure.
Thus, each $F \in \Rdists$ lies within a subset of $\Rdists$ on which $\leq_3$ is transitive. In light of these observations, one could adapt the classical order-based approach to define measures of location, dispersion and skewness to kurtosis. Instead of requiring a mapping $\kappa: \Rdists \to \R$ to generally preserve the order $\leq_3$, one could require the restriction of $\kappa$ to the transitivity set $\mathcal{T}_{D,1/2}^t$ to preserve $\leq_3$ for all $t \in \R$.
%Alternatively, one could also choose a $p \in (0, 1) \setminus \{1/2\}$.

These observations raise the question whether there exist other skewness measures that induce transitivity sets analogous to Theorem \ref{thm:kurtDensTrans}. To that end, note that a simple sufficient condition for the term $\gamma_D^{p_0}(F)$ to coincide is to require $f'(F^{-1}(p_0)) = 0$ for all cdf's $F$ in the given set.
Hence, for each $p_0 \in (0, 1)$, $\leq_3$ is transitive on the set of all cdf's, the density of which has a stationary point at the $p_0$-quantile. One well known point, at which this commonly occurs, is the mode of a distribution.
For the following considerations, we assume that all distributions are unimodal and denote the mode of $F$ by $M_F$. If the mode lies in the interior of the support, the assumptions on $F$ directly yield $f'(M_F) = 0$. It follows that, for any $p \in (0, 1)$, $\gamma_D^p(F) = 0$ holds for all cdf's $F$ in the set
\begin{equation*}
%\label{eqn:defModeTransSets}
	\mathcal{T}_{Mode}^{\tilde{p}} = \{F: M_F = F^{-1}(p)\} = \{F: 1 - 2 F(M_F) = \tilde{p}\},
\end{equation*}
where $\tilde{p} = 1 - 2p$.
In combination with Propositions \ref{thm:sameIV_leq3Trans} and \ref{thm:constDensSkew_leq3Trans}, this observation  yields the following result.

\begin{theorem}
\label{thm:leq3TransMode}
	For any $\tilde{p} \in (-1, 1)$, the kurtosis order $\leq_3$ is transitive on the set $\mathcal{T}_{Mode}^{\tilde{p}}$.
\end{theorem}

For any $\tilde{p} \in (-1, 1)$ and any pair of cdf's $F, G \in \mathcal{T}_{Mode}^{\tilde{p}}$ with $F \leq_3 G$, the corresponding inflection value is given by $p = (\tilde{p}+1)/2$. \citet[p.\ 35]{ag_mode} showed that $\gamma_{Mode}(F) = 1 - 2 F(M_F), F \in \Rdists,$ is a measure of skewness, which entails that it preserves the skewness order $\leq_2$. Thus, the transitivity of $\leq_3$ on the sets $\mathcal{T}_{Mode}^{\tilde{p}}$ has a similar interpretation to before: for $\leq_3$ to be transitive, the skewness of the involved distributions needs to be constant in some sense.

For distributions with modes at the boundaries of their supports, the above transitivity property does not hold, i.e., $\leq_3$ is not transitive on $\mathcal{T}_{Mode}^{-1}$ and $\mathcal{T}_{Mode}^1$ in general. The crucial result in Proposition \ref{thm:constDensSkew_leq3Trans} does not hold in these cases. Counterexamples can be constructed using Weibull distributions, applying the results given in Section \ref{sec:dists} below. Thus, the sets $\mathcal{T}_{Mode}^{\tilde{p}}, \tilde{p} \in (-1, 1),$ do not provide a partition of the set of all (sufficiently regular) probability distributions on the real numbers.
%as the ones obtained for the sets $\mathcal{T}_{D, p}^t$.

The notion of a mode can be generalized without losing the transitivity of $\leq_3$ on the corresponding sets $\mathcal{T}_{Mode}^{\tilde{p}}$. Specifically, Theorem \ref{thm:leq3TransMode} still holds if $f$ only attains a local maximum at $M_F$, no longer assuming $F$ to be unimodal. However, \citet{ag_mode} only proved $\gamma_{Mode}$ to be a skewness measure under the assumption of unimodality.

The relationships between the transitivity sets found in this section and their connection to the set of all symmetric distributions are summarized in the following remark.

\begin{remark}
\label{rmk:symmTransSetLeq3}
	Let $\tilde{p} \in (-1, 1)$ and let $F \in \mathcal{T}_{Mode}^{\tilde{p}}$ be unimodal. It follows that $M_F = F^{-1}(p)$, where $p = (\tilde{p}+1)/2 \in (0, 1)$. Since $M_F$ lies within the interior of the support of $F$, we obtain $f'(F^{-1}(p)) = 0$ and therefore $\gamma_D^{p}(F) = 0$. Thus, the inclusion $\mathcal{T}_{Mode}^{\tilde{p}} \subseteq \mathcal{T}_{D, p}^0$ holds for all $p \in (0, 1)$ with $\tilde{p} = 2p-1$. In particular, $\mathcal{T}_{Mode}^{0} \subseteq \mathcal{T}_{D, 1/2}^0$.
	
	Now, let $F \in \Rdists$ be symmetric, denoted by $F \in \Sdists$. Since both $\gamma_D^p$ and $\gamma_{Mode}$ are invariant under transformations of the form $x \mapsto a x + b$ for $a > 0$ and $b \in \R$, we can assume without restriction that the symmetry centre of $F$ is $0$. Because this implies $\gamma_D^{1/2}(F) = f'(0)/(f(0))^2 = 0$, we obtain the inclusion $\Sdists \subseteq \mathcal{T}_{D, 1/2}^0$. If, additionally, $F$ is assumed to be unimodal, $M_F = 0$ and $\gamma_{Mode}(F) = 0$ follows. Thus, in this case,
	$\mathcal{S} \subseteq \mathcal{T}_{Mode}^0 \subseteq \mathcal{T}_{D,1/2}^0$ holds.
\end{remark}

Since $\leq_3$ is transitive on $\mathcal{T}_D^0$, it is also transitive on the set of all symmetric cdf's. \citet{oja}, virtually the only work which mentions the order $\leq_3$, dismissed it due to its non-transitivity, and instead focused on the previously mentioned concave-convex order $\leq_s$. However, Oja restricted his considerations concerning kurtosis to symmetric distributions, and therefore also proved the transitivity of $\leq_s$ only on this class. Since $\leq_3$ is also transitive on symmetric distributions, Oja's argument is not convincing.
%\citet{oja}, as the only paper with a significant mention of the order $\leq_3$ in the literature, dismissed it and instead focussed on the previously mentioned concave-convex order $\leq_s$. The only given reason is that $\leq_3$ is not transitive. Oja, however, restricted all of his considerations concerning kurtosis to symmetric distributions and therefore also only proved the transitivity of $\leq_s$ on this class. Since $\leq_3$ is also transitive on symmetric distributions, Oja's argument for dismissing it is refuted.}}

\subsection{Equivalence with respect to $\boldsymbol{\leq_3$}}

Two distributions $F, G \in \Rdists$ are said to be equivalent with respect to $\leq_3$, denoted by $F =_3 G$, if both $F \leq_3 G$ and $G \leq_3 F$ hold. This is equivalent to $R_{FG}''' \geq 0$ and $R_{GF}''' \geq 0$. Using $R_{GF}=R_{FG}^{-1}$ to rewrite the third derivative of $R_{GF}$ as
\begin{align}
% \nonumber
% 	R_{GF}'(t) &= \frac{1}{R_{FG}'(R_{GF}(t))},\\
% \label{eqn:scndAblRGF}
% 	R_{GF}''(t) &= -\frac{R_{FG}''(R_{GF}(t))}{(R_{FG}'(R_{GF}(t)))^3},\\
% \nonumber
% 	R_{GF}'''(t) &= 3 \frac{(R_{FG}''(R_{GF}(t)))^2}{(R_{FG}'(R_{GF}(t)))^5} - \frac{R_{FG}'''(R_{GF}(t))}{(R_{FG}'(R_{GF}(t)))^4}\\
\nonumber
	R_{GF}'''(t) &= \frac{3 (R_{FG}''(R_{GF}(t)))^2 - R_{FG}'''(R_{GF}(t)) R_{FG}'(R_{GF}(t))}{(R_{FG}'(R_{GF}(t)))^5},
\end{align}
it follows that
\begin{align*}
	G \leq_3 F
% 	&\Leftrightarrow R_{GF}'''(t) \geq 0 \quad \forall t \in D_G\\
% 	&\Leftrightarrow \frac{3 (R_{FG}''(t))^2 - R_{FG}'''(t) \cdot R_{FG}'(t)}{(R_{FG}'(t))^5} \geq 0 \quad \forall t \in D_F\\
	&\Leftrightarrow R_{FG}'''(t) \leq 3 \frac{(R_{FG}''(t))^2}{R_{FG}'(t)} \quad \forall t \in D_F.
\end{align*}
% Thus, $F =_3 G$ is equivalent to
% \begin{equation*}
% 	0 \leq R_{FG}'''(t) \leq 3 \frac{(R_{FG}''(t))^2}{R_{FG}'(t)}
% \end{equation*}
% for all $t \in D_F$. Obviously, an analogous condition applies to $R_{GF}$.
Hence, we have the following result.

\begin{proposition}
\label{thm:=_3Equiv}
	$F =_3 G$ holds, if and only if $R_{FG}$ satisfies the differential inequality
	\begin{equation}
	\label{eqn:=_3DUGL}
		0 \leq \varphi'''(t) \leq 3 \frac{(\varphi''(t))^2}{\varphi'(t)} \quad \forall t \in D_F.
	\end{equation}
\end{proposition}

The fact that $F =_3 G$ is not equivalent to $R_{FG}''' \equiv 0$ is notable as it systematically differs from what can be observed with the orders $\leq_0$, $\leq_1$ and $\leq_2$ of location, dispersion and skewness. Equivalence with respect to any of these orders occurs if and only if the corresponding derivative of $R_{FG}$ is constantly zero. Thus, $F =_0 G$ is equivalent to $F = G$, $F =_1 G$ is equivalent to $F(\cdot) = G(\cdot + b)$ for a $b \in \R$, and $F =_2 G$ is equivalent to $F(\cdot) = G(a \cdot + b)$ for an $a > 0$ and a $b \in \R$. Heuristically, equivalence with respect to dispersion means that $G$ is a shifted or relocated version of $F$ and equivalence with respect to skewness means that $G$ is a shifted and rescaled version of $F$, allowing for changes in location and dispersion. This suggests that the functions satisfying the differential inequality (\ref{eqn:=_3DUGL}) can change the location, the dispersion and the skewness of a distribution while being kurtosis-invariant. However, the fact that this family of functions are not as simple as the family of all affine linear transformations suggests that there exists no simple operation to standardize distributions with respect to skewness.

In the following example, Proposition \ref{thm:=_3Equiv} is applied to monomials.

\begin{example}
\label{exm:=_3Monom}
	Let $R_{FG}(t)=t^p, 0<t<1,$ for some $p>0$. This arises, for example, for $F(t)=t, G(t)=t^{1/p}$, for $F(t)=t^p, G(t)=t$, or, with support $t>0$, for Weibull distributions (see Section \ref{sec:dists}). For $p \notin \{1, 2\}$, $F \leq_3 G$ is equivalent to
	\begin{equation*}
		0 \leq R_{FG}'''(t) = p(p-1)(p-2)t^{p-3} \ \forall t
		\quad \Leftrightarrow \quad p \notin (1, 2).
	\end{equation*}
	Since $R_{FG}''' \equiv 0$ for $p \in \{1, 2\}$, $F \leq_3 G$ is equivalent to $p \notin (1, 2)$. Conversely, for $p \notin \{1, 2\}$, $G \leq_3 F$ is equivalent to
	\begin{align*}
		R_{FG}'''(t) \leq 3 \frac{(R_{FG}(t)'')^2}{R_{FG}'(t)} = 3p(p-1)^2 t^{p-3} \ \forall t
		\quad \Leftrightarrow \quad p \notin (\tfrac{1}{2}, 1).
	\end{align*}
	Since the inequality is obviously satisfied for $p \in \{1, 2\}$, $G \leq_3 F$ is equivalent to $p \notin (\frac{1}{2}, 1)$. Overall, $F =_3 G$ is satisfied, if and only if
	$$
	p \in (0,1/2] \cup \{1\} \cup [2,\infty).
	$$
	In particular, $F(t)=t, t \in (0, 1),$ and $G(t)=t^2, t \in (0, 1),$ are equivalent with respect to $\leq_3$.
	
	%Clearly, in this example, one can get the result also by checking $R_{FG}'''(t)\ge 0$ and $R_{GF}'''(t)\ge 0$ for $t\in(0,1)$, but the advantage of using Proposition \ref{prop-3-equal} in general is that only the derivatives of one of the two functions must be known.
\end{example}

Note that $R_{FG}''' \equiv 0$ and therefore also $F =_3 G$ holds if $R_{FG}$ is any polynomial of degree $\leq 2$.
%This result can also easily be obtained using Corollary \ref{thm:kCxCharIntersect}.
While $F =_2 G$ is equivalent to $R_{FG}$ being a polynomial of degree $\leq 1$, the fact that $R_{FG}$ is a polynomial of degree $\leq 2$ is only a sufficient, but not a necessary condition for $F =_3 G$.\par
%The result of Example \ref{exm:=_3Monom} can also be obtained by checking $R_{FG}'''(t)\ge 0$ and $R_{GF}'''(t)\ge 0$ for $t\in(0,1)$, but the advantage of using Proposition \ref{thm:=_3Equiv} is that only the derivatives of one of the two functions must be known. This is illustrated by the following example.

\section{Concave-convex kurtosis orders}
\label{sec:leqgs}

In the literature, there exist two major proposals for generalizing the concave-convex order $\leq_s$ to asymmetric distributions, denoted by $\leq_a$ and $\leq_S$ (see \citealp{MacG_kurt_skew} and \citealp{MacG_kurt_spread}). The order $\leq_S$ is not considered further in the present work because it disregards a critical amount of information, as expanded upon in Section \ref{sec:intro}.
The critical drawback of the order $\leq_a$ can best be explained using the notion of the inflection value from Definition \ref{def:inflVal}. Just like in our considerations in Section \ref{sec:leq3TS}, $F \leq_a G$ requires that the function $R_{FG}$ has one change from negative to positive curvature, whose location can be identified by an inflection value $p_{FG} \in (0, 1)$. While $p_{FG} = 1/2$ necessarily holds if $F$ and $G$ are symmetric, there is no reason to assume it to be a prerequisite for two asymmetric distributions to be ordered with respect to kurtosis. Thus, whereas the generally applicable order $\leq_3$ is stronger than $\leq_s$ in a symmetric setting, the same can not be said about the generalized version $\leq_a$ of $\leq_s$ in a general setting. In the following, we propose an alternative generalization of $\leq_s$ that is not a priori restricted to a specific inflection value.

\begin{definition}
\label{def:leqgs}
	$F$ is said to be {\em less kurtotic in the concave-convex sense} than $G$, denoted by $F \leq_{gs} G$, if there exists a $p_{FG} \in [0, 1]$ such that $R_{FG}$ is concave on $D_F \cap (- \infty, F^{-1}(p_{FG}))$ and convex on $D_F \cap (F^{-1}(p_{FG}), \infty)$.
\end{definition}

The fact that $F \leq_3 G$ implies $F \leq_{gs} G$ for all $F, G \in \Rdists$ is a direct consequence of Theorem \ref{thm:leq3leqgst0} below. The essential difference between the two orders is that the first requires that a function (in this case $R_{FG}''$) is increasing whereas the second requires that the same function changes values from negative to positive at some point. This principle has also been used in the literature to obtain weakenings of other orders from the family $\leq_k, k \in \N_0$.
As an example, we can consider the visually more striking characteristic of dispersion based on the order $\leq_1$. Instead of assuming that $\Delta_{FG}$ increases, which is equivalent to $F \leq_1 G$, we can require that the values of $\Delta_{FG}$ switch from negative to positive at some point. A similar dispersion order has been proposed by \citet[p.\ 158]{oja}. He writes $F \leq_1^* G$ if there exists $x_0 \in D_F$ such that $\Delta_{FG}(x) \leq E(Y) - E(X)$ for $x \leq x_0$ and $\Delta_{FG}(x) \geq E(Y) - E(X)$ for $x \geq x_0$. The sole difference to the order introduced before is the threshold, which changes from zero to the difference of the expectations. Unlike zero, the difference of the expectations is guaranteed to be taken as a value of $\Delta_{FG}$ at some point. This can be seen by considering the centred versions of $F$ and $G$. If, for example, the locations of $F$ and $G$ differ substantially, using the threshold zero is obviously not reasonable.

This line of arguments can also be applied to the order $\leq_{gs}$ and the function $R_{FG}''$. For general distribution functions $F$ and $G$, there is no reason to assume that $R_{FG}''$ takes the value zero at some point. Thus, Definition \ref{def:leqgs} needs to be modified. However, because $F$ and $G$ can only be standardized with respect to location and dispersion and not with respect to skewness, we cannot use the same technique as for $\leq_1^*$ to obtain an alternative threshold. Therefore, the following definition uses a variable threshold.

\begin{definition}
\label{def:leqgst_0}
	Let $t_0 \in \R$. Then, $F$ is said to be {\em less kurtotic than $G$ in the concave-convex sense with threshold $t_0$}, denoted by $F \leq_{gs}^{t_0} G$, if there exists a $p_{FG}^{t_0} \in [0, 1]$ such that $R_{FG}''(t) \leq t_0$ holds for all $t \in D_F \cap (- \infty, F^{-1}(p_{FG}^{t_0}))$ and $R_{FG}''(t) \geq t_0$ holds for all $t \in D_F \cap (F^{-1}(p_{FG}^{t_0}), \infty)$.
\end{definition}

Note that the orders $\leq_{gs}^{0}$ and $\leq_{gs}$ coincide. While the order $\leq_{gs}^{t_0}$ is formally defined for all $t_0 \in \R$, it is only meaningful if $t_0 \in \mathrm{int}(R_{FG}''(D_F))$. Otherwise, it is obvious that either $R_{FG}''(t) \leq t_0$ or $R_{FG}''(t) \geq t_0$ holds for all $t \in D_F$. Hence, all thresholds $t_0 \in \mathrm{int}(R_{FG}''(D_F))$ are said to be {\em reasonable}. The only exception is the case that the set of reasonable thresholds is empty, which is equivalent to $R_{FG}''$ being constant. In this case, the sole value of $R_{FG}''$ is the only candidate for a reasonable threshold.

The relationship between $\leq_3$ and the family $\leq_{gs}^{t_0}, t_0 \in \R,$ given in the following theorem, underpins the idea that the latter consists of natural weakenings of $\leq_3$.

\begin{theorem}
\label{thm:leq3leqgst0}
	Let $F, G \in \Rdists$. Then, $F \leq_3 G$ is equivalent to $F \leq_{gs}^{t_0} G$ for all $t_0 \in \mathrm{int}(R_{FG}''(D_F))$.
%The set of values for $t_0$ can be extended to $\R$ without losing the equivalence.
	
	\begin{proof}
	    The implication from left to right holds by construction. For the reverse implication, let $t_1 \in D_F$. If $t_1$ lies within an interval on which $R_{FG}''$ is constant, $R_{FG}'''(t_1) = 0$ follows. Otherwise, it follows that $t_0 = R_{FG}''(t_1) \in \mathrm{int}(R_{FG}''(D_F))$. Now
		\begin{equation*}
			R_{FG}'''(t_1) = \lim_{\varepsilon \searrow 0} \frac{R_{FG}''(t_1 + \varepsilon) - R_{FG}''(t_1 - \varepsilon)}{2 \varepsilon} \geq 0
		\end{equation*}
		holds because of $R_{FG}''(t_1 + \varepsilon) \geq t_0$ and $R_{FG}''(t_1 - \varepsilon) \leq t_0$ by assumption. The assertion follows since $t_1$ was arbitrary. %\par
		%The extension of the set of values for $t_0$ to $\R$ is valid because either $R_{FG}''(t) \leq t_0$ or $R_{FG}'' \geq t_0$ is true by construction for all unreasonable thresholds $t_0 \notin \mathrm{int}(R_{FG}''(D_F))$.
	\end{proof}
\end{theorem}

In Theorem \ref{thm:leq3leqgst0}, the set $\mathrm{int}(R_{FG}''(D_F))$ can be replaced by  $\R$ because either $R_{FG}''(t) \leq t_0$ or $R_{FG}'' \geq t_0$ is true by construction for all unreasonable thresholds $t_0 \notin \mathrm{int}(R_{FG}''(D_F))$.

The following result states that the proposed extension of the concave-convex order $\leq_s$ to asymmetric distributions is not transitive in general, implying that it is not superior to $\leq_3$ in this respect.

\begin{proposition}
\label{thm:trans_leqgs}
	For all $t_0 \in \R$, the kurtosis order $\leq_{gs}^{t_0}$ is not transitive in general.
	%The specific strict order $<_{gs}^0$, however, is transitive on the sets $\mathcal{T}_{Mode}^{\tilde{p}}, \ \tilde{p} \in (-1, 1),$ and the sets $\mathcal{T}_D^t, \ t \in \R$.
	
	\begin{proof}
		A counterexample can be obtained for all $t_0 \in \R$ by reusing Example \ref{exm:CounterTransLeq3} with a rescaled version of $H^{-1}$. For that, let  $c>0$ and
		\begin{equation*}
			H: [0, c] \to [0, 1], \quad t \mapsto 1 - \sqrt[3]{\frac{c-t}{c}}.
		\end{equation*}
		This implies that the functions $R_{GH}$ and $R_{FH}$ as well as all of their derivatives are multiplied by the factor $c$. So, additionally to $F \leq_3 G$, $R_{GH}'''(t) = 6c \geq 0$ holds for all $t \in [0, 1]$, and, thus, $G \leq_3 H$. By Theorem \ref{thm:leq3leqgst0}, $F \leq_{gs}^{t_0} G$ and $G \leq_{gs}^{t_0} H$ hold for all $t_0 \in \R$. In contrast, we have
		\begin{equation*}
			R_{FH}''(t) = 18c (4t^7-5t^4+t)
			\begin{cases}
				< 0 \quad &\text{ for } t \in (2^{-\frac{2}{3}}, 1),\\
				= 0 \quad &\text{ for } t \in \{0, 2^{-\frac{2}{3}}, 1\},\\
				> 0 \quad &\text{ for } t \in (0, 2^{-\frac{2}{3}}).
			\end{cases}
		\end{equation*}
		It follows that, for any $t_0 > 0$, there exists $c > 0$ such that $R_{FH}''$ first takes values smaller than $t_0$, then larger, and finally smaller again. For any $t_0 < 0$, there exists $c > 0$ such that $R_{FH}''$ first takes values larger than $t_0$, then smaller and finally larger again. For $t_0 = 0$, we obtain $R_{FH}''(t) \geq 0$ for $t \leq 2^{-2/3}$ and $R_{FH}''(t) \leq 0$ for $t \geq 2^{-2/3}$. All three cases pose a contradiction to $F \leq_{gs}^{t_0} G$.
		%Let $\tilde{p} \in (-1, 1)$ and  $F, G, H \in \mathcal{T}_{Mode}^{\tilde{p}}$ with $F <_{gs} G$ and $G <_{gs} H$. By the line of reasoning used in the proof of Theorem \ref{thm:KurtModeTrans}, $\Delta_{FG}''(F^{-1}(p)) = 0 = \Delta_{GH}''(G^{-1}(p))$, where $p = \frac{\tilde{p}+1}{2}$. Since, by definition of $<_{gs}$, there exists at most one $t \in \supp(F)$ and one $s \in \supp(G)$ such that $\Delta_{FG}''(t) = 0$ and $\Delta_{GH}''(s) = 0$, $t = F^{-1}(p)$ and $s = G^{-1}(p)$ follows. Considering (\ref{eqn:2ndDerivativeDeltaFH}) along with the fact that $\Delta_{GH}$ is increasing, this yields $\Delta_{FH}''(F^{-1}(q)) < 0$ for $q < p$ and $\Delta_{FH}''(F^{-1}(q)) > 0$ for $q > p$. Overall, $F <_{gs} H$.\par
		%The transitivity on the sets $\mathcal{T}_D^t, \ t \in \R,$ can be proved analogously and only differs in the way that the second derivatives of $\Delta_{FG}$ and $\Delta_{GH}$ are shown to be zero in $F^{-1}(p)$ and $G^{-1}(p)$, respectively.
	\end{proof}
\end{proposition}

For symmetric cdf's $F$ and $G$, $R_{FG}$ always has an inflection point at $F^{-1}(1/2)$. Thus, $\leq_{gs}$ is equivalent to $\leq_s$ on $\Sdists$ and therefore also transitive on $\Sdists$ \citep[see][p.\ 165]{oja}. The situation is different for $\leq_{gs}^{t_0}, t_0 \neq 0$ because the critical switch from $R_{FG}''(t) \leq t_0$ to $R_{FG}''(t) \geq t_0$ cannot occur at $F^{-1}(1/2)$ due to the point symmetry of $R_{FG}$.

\begin{remark}
\label{rmk:leqgsTrans}
    The specific order $\leq_{gs}^0$ (or, equivalently, $\leq_{gs}$) can be altered slightly to become transitive on the more general sets $\mathcal{T}_{Mode}^{\tilde{p}}, \tilde{p} \in (-1, 1),$ and $\mathcal{T}_{D, p}^t, t \in \R, p \in (0, 1)$. For two cdf's $F$ and $G$, we say that $F <_{gss} G$ holds if there exists a $p_{FG} \in [0, 1]$ such that $R_{FG}''$ is strictly negative on $D_F \cap (- \infty, F^{-1}(p_{FG}))$, and strictly positive on $D_F \cap (F^{-1}(p_{FG}), \infty)$. Note that $<_{gss}$ is not equivalent to $<_{gs}$ since the latter is defined by
	\begin{equation*}
		F <_{gs} G \Leftrightarrow F \leq_{gs} G \text{ and } F \neq_{gs} G \Leftrightarrow F \leq_{gs} G \text{ and } G \not\leq_{gs} F,
	\end{equation*}
	as usual for strict versions of orders. To see that $<_{gss}$ is transitive, let $p \in (0, 1)$ and  $F, G, H \in \mathcal{T}_{D, p}^{t}$ with $F <_{gss} G$ and $G <_{gss} H$. By the line of reasoning used to prove Proposition \ref{thm:constDensSkew_leq3Trans} and Theorem \ref{thm:kurtDensTrans}, $R_{FG}''(F^{-1}(p)) = 0 = R_{GH}''(G^{-1}(p))$ then holds. Since, by definition of $<_{gss}$, there exists at most one $t \in D_F$ and one $s \in D_G$ such that $R_{FG}''(t) = 0$ and $R_{GH}''(s) = 0$, $t = F^{-1}(p)$ and $s = G^{-1}(p)$ follows. Considering (\ref{eqn:2ndDerivativeDeltaFH}) for $t = F^{-1}(p)$ along with the fact that $R_{GH}$ is increasing, this yields $R_{FH}''(F^{-1}(q)) < 0$ for $q < p$ and $R_{FH}''(F^{-1}(q)) > 0$ for $q > p$. Overall, $F <_{gss} H$ follows.
	The transitivity of $<_{gss}$ on the sets $\mathcal{T}_{Mode}^{\tilde{p}}, p \in (-1, 1),$ now follows from $\mathcal{T}_{Mode}^{\tilde{p}} \subseteq \mathcal{T}_{D, p}^0$, where $p = (\tilde{p}+1)/2$.
\end{remark}

It is not possible to show the transitivity of the order $\leq_{gs}$ on the given sets in the same way as for $<_{gss}$, since, assuming $F \leq_{gs} G$, $R_{FG}''(F^{-1}(p)) = 0$ for any $p \in (0,1)$ is not sufficient to infer that $p$ is an inflection value.
Because the concavity and the convexity of $R_{FG}$ on either side of the actual inflection value is not assumed to be strict, the function could be convex on both sides of $F^{-1}(p)$ or concave on both sides.

\section{Application to specific distributions}
\label{sec:dists}

\subsection{Weibull distribution}

As an example of a well-known family of distributions with varying degrees of skewness, we consider Weibull distributions. Without restriction, we set the scale parameter to 1, and denote the distribution with shape parameter $k$ by $\W(k)$. Let $X \sim \W(k), Y \sim \W(\ell)$ for $0<k<\ell$.
For $t > 0$, we have $R_{FG}(t) = t^{k/\ell}$. It follows directly from Example \ref{exm:=_3Monom} that $F \leq_3 G$ holds for all $k<\ell$, whereas $F =_3 G$ holds for $2k\leq \ell$. Thus, if the two parameters differ by less than a factor two, the distribution with the higher parameter value is strictly more kurtotic. If the two parameters differ at least by a factor two, the two distributions are equivalent with respect to the order $\leq_3$. Considering that a large difference between the two parameter values is also associated with a large difference in skewness, this may best be interpreted as follows. If the difference in skewness between two Weibull distributions is too large, they cannot be unambiguously ordered with respect to kurtosis.

This rather unintuitive behaviour allows us to construct another counterexample for the transitivity of $\leq_3$ since, e.g.,
$\W(k) \leq_3 \W(1.5k) =_3 \W(0.7k) \not\geq_3 \W(k)$
holds for all $k > 0$.
Furthermore, it is easy to show that $\leq_{gs}^{t_0}$ coincides with $\leq_3$ on the family of Weibull distributions for all reasonable thresholds $t_0$. Thus, the given counterexample also applies to $\leq_{gs}^{t_0}$.

\subsection{Sinh-arsinh distribution}

The family of sinh-arsinh distributions was introduced by \citet{Jones:2009}. It is dependent upon four parameters, which are associated with location, dispersion, skewness and tailweight. Here, we consider a simplified two-parameter family by fixing the location and dispersion parameters to zero and one, respectively. A random variable $X$ is said to be sinh-arsinh-distributed with skewness parameter $\nu \in \R$ and tailweight $\tau > 0$, denoted by $X \sim \SAS(\nu, \tau)$, if the random variable
\begin{equation*}
	Z = S_{\nu, \tau}(X) = \sinh(\tau \cdot \operatorname{arsinh}(X) - \nu)
\end{equation*}
is standard normal. Skewness to the right increases with increasing $\nu$ and tailweight decreases with increasing $\tau$. More specifically, $F \leq_2 G$ if $\nu_F \leq \nu_G, \tau_F = \tau_G$ and $F \leq_{gs} G$ if $\nu_F = \nu_G = 0, \tau_F \leq \tau_G$ \citep[see][pp.\ 763, 765, 766]{Jones:2009}. One can directly infer the corresponding distribution function $F = \Phi \circ S_{\nu, \tau}$ and quantile function $F^{-1} = S_{\nu, \tau}^{-1} \circ \Phi^{-1} = S_{-\nu/\tau, 1/\tau} \circ \Phi^{-1}$ of $X$.

There exist numerous other distribution families with four parameters that are associated with location, dispersion, skewness and tailweight or kurtosis. Examples include the skew-$t$ distribution \citep{skew-normal,skew-t} and Tukey's $g$-and-$h$ or $g$-and-$k$ distributions \citep{tukey-g-and-h,g-and-h_hoaglin,g-and-k}. However, these families do not have similarly explicit representations of both their distribution and quantile functions. Furthermore, while the skew-$t$ distributions do include the standard normal distribution, it only appears as a limiting case and not as a standard case as for the sinh-arsinh distributions. Finally, the sinh-arsinh transformation can also be applied to (symmetric) base distributions other than the standard normal. For example, \citet{sinh-arcsinh-t} applied it to Student's $t$-distribution.

Let $X \sim \SAS(\nu_F, \tau_F)$ and $Y \sim \SAS(\nu_G, \tau_G)$ with distribution functions $F$ and $G$. It follows that $R_{FG}(t) = S_{\tilde{\nu}, \tilde{\tau}}(t)$, where $\tilde{\nu} = (\nu_F - \nu_G)/\tau_G$ and $\tilde{\tau} = \tau_F/\tau_G$. Note that the fulfilment of $F \leq_{gs}^{t_0} G$ and $F \leq_3 G$ is solely dependent on $R_{FG}$. Hence, the ordering of $F$ and $G$ in terms of kurtosis only depends upon two parameters instead of four.
%Particularly, it is independent of the concrete values of the skewness parameters and instead only depends on their difference.
%
The following result gives conditions for the ordering of sinh-arsinh distributions with respect to the kurtosis orders $\leq_3$ and $\leq_{gs}$.

\begin{theorem}
\label{thm:SASKurtOrd}
	Let $F \neq G$  and $t_0 \in \mathrm{int}(R_{FG}''(D_F))$. Then, $F \leq_3 G$ holds if and only if $\tau_F \geq 2 \tau_G$.
	Likewise, for $t_0\neq 0$, $F \leq_{gs}^{t_0} G$ is equivalent to $\tau_F \geq 2 \tau_G$.
    Furthermore, $F \leq_{gs}^0 G$ if and only if $\tau_F > \tau_G$.
\end{theorem}

The key characteristics of $R_{FG}''$ are summarized in Table \ref{tbl:KurtOrdSAS}. The proof of Theorem \ref{thm:SASKurtOrd} can be found in the appendix. %supplementary material.

%\begin{remark}
%\textcolor{red}{ins Supplement nach Beweis}
%	By examining the proof of Theorem \ref{thm:SASKurtOrd}, it can be seen that $F \neq G$ implies $\mathrm{int}(R_{FG}''(D_F)) \neq \emptyset$, so that parts b) and c) of Theorem \ref{thm:SASKurtOrd} are not statements concerning the empty set.
%\end{remark}

Since the usual order of the real numbers used in the equivalent conditions in Theorem \ref{thm:SASKurtOrd} is transitive, the following result is directly implied.

\begin{corollary}
\label{thm:leq3TransSAS}
	Let $t_0 \in \mathrm{int}(R_{FG}''(D_F))$. Then, the orders $\leq_3$ and $\leq_{gs}^{t_0}$ are transitive on the set $\{F \in \Rdists: \exists \nu \in \R, \tau > 0: F = \SAS(\nu, \tau)\}$.
\end{corollary}

\begin{table}
	\centering
	\begin{tabular}{c||c|c|c|c|c}
		& \multicolumn{3}{c|}{$R_{FG}''$} &&\\
		\hline
		Value of $\tilde{\tau}$ & Sign change? & Monotonicity? & $\lim_{t \to \pm \infty}$ & $F \leq_3 G$? & $F \leq_{gs}^{t_0} G$?\\
		\hline\hline
		$\tilde{\tau} \in (0, 1)$ & '$+$' to '$-$' & No & $0$ & No & No\\
		\hline
		$\tilde{\tau} = 1$ & No & No* & $0$ & No** & No**\\
		\hline
		$\tilde{\tau} \in (1, 2)$ & '$-$' to '$+$' & No & $0$ & No & Iff $t_0 = 0$\\
		\hline
		$\tilde{\tau} = 2$ & '$-$' to '$+$' & Increasing & $\pm 4$ & Yes & Yes\\
		\hline
		$\tilde{\tau} \in (2, 3)$ & '$-$' to '$+$' & Increasing & \begin{tabular}{@{}c@{}}$\pm \infty$ \\ sub-linear growth\end{tabular} & Yes & Yes\\
		\hline
		$\tilde{\tau} = 3$ & '$-$' to '$+$' & Increasing & \begin{tabular}{@{}c@{}}$\pm \infty$ \\ linear growth\end{tabular} & Yes & Yes\\
		\hline
		$\tilde{\tau} \in (3, \infty)$ & '$-$' to '$+$' & Increasing & \begin{tabular}{@{}c@{}}$\pm \infty$ \\ super-linear growth\end{tabular} & Yes & Yes
	\end{tabular}
	\caption{\label{tbl:KurtOrdSAS} Behaviour of the function $R_{FG}''$ and kurtosis orders for distribution functions $F$ and $G$ of $X \sim \SAS(\nu_F, \tau_F)$ and $Y \sim \SAS(\nu_G, \tau_G)$. \
		*: Constant, if $\tilde{\nu} = 0$.
		**: Yes, if $\tilde{\nu} = 0$.}
\end{table}

Heuristically, Theorem \ref{thm:SASKurtOrd} implies that, within the family of sinh-arsinh distributions, comparisons in terms of kurtosis are skewness-invariant. This is due to the fact that equivalent characterizations for both major kurtosis orders are independent of both $\nu_F$ and $\nu_G$, which are skewness parameters by construction and also in the sense of $\leq_2$ for $\tau_F = \tau_G$ \citep[see][p.\ 763]{Jones:2009}. Moreover, the characterizations in Theorem \ref{thm:SASKurtOrd} not only stay the same for equally skewed asymmetric distributions, but also for pairs of distributions with arbitrarily big differences in skewness. Also note that these results can be generalized to families of sinh-arsinh distributions that arise from symmetric base distributions other than the normal since the functions $R_{FG}$ only depend on the transformations and not on the specific base distribution.
%This observation serves as a strong argument for considering the notion of kurtosis irrespective of skewness, in particular using the two kurtosis orders used in Theorem \ref{thm:SASKurtOrd}.\par
The skewness-invariance of the sinh-arsinh distribution in terms of kurtosis was noted by \citet[pp.\ 91--92]{jones_quint}. Specifically, they showed that quantile-based kurtosis measures that are constructed from symmetric differences of the form $F^{-1}(1-\alpha) - F^{-1}(\alpha), \alpha \in (0, 1/2),$ are invariant under changes of the skewness parameter $\nu$. Theorem \ref{thm:SASKurtOrd} generalizes this skewness-invariance from a specific family of kurtosis measures to the underlying kurtosis orders.

%\section{Discussion}

% \section*{Acknowledgement}
% Acknowledgements should appear after the body of the paper but before any appendices and be as brief as possible
% subject to politeness. Information, such as contract numbers, of no interest to readers, must
% be excluded.

\section*{Supplementary material}
\label{SM}
%Supplementary material available at \Bka\ online
The appendix includes the proof of Theorem \ref{thm:SASKurtOrd} along with some illustrations of the functions described in Table \ref{tbl:KurtOrdSAS}.

\vspace*{-10pt}

\setlength{\bibsep}{3pt}
%\bibliographystyle{biometrika.bst}
%\bibliography{paper-ref}

\appendix

\section{Proof of Theorem \ref{thm:SASKurtOrd}}

In order to prove the equivalent characterization of $F \leq_3 G$ and that of $F \leq_{gs}^{t_0} G$ for $t_0 \neq 0$, we assume $t_0 \in \mathrm{int}(R_{FG}''(D_F)) \setminus \{0\}$ and prove the chain of implications
\begin{equation}
\label{eqn:leq3gsSASChain}
    F \leq_{gs}^{t_0} G \Rightarrow \tau_F \geq 2\tau_G \Rightarrow F \leq_3 G.
\end{equation}
Since $F \leq_3 G$ implies $F \leq_{gs}^{t_0} G$ due to Theorem 3 in the main paper, all three statements are then equivalent. Because of $R_{FG}(t) = S_{\tilde{\nu}, \tilde{\tau}}(t)$, where $\tilde{\nu} = (\nu_F - \nu_G)/\tau_G$ and $\tilde{\tau} = \tau_F/\tau_G$, it follows that
\begin{align*}
	R_{FG}'(t) &= \tilde{\tau} (1 + t^2)^{-1/2} C_{\tilde{\nu}, \tilde{\tau}}(t),\\
	R_{FG}''(t) &= \tilde{\tau}(1 + t^2)^{-3/2} \left[ \tilde{\tau} \sqrt{1 + t^2} S_{\tilde{\nu}, \tilde{\tau}}(t) - t C_{\tilde{\nu}, \tilde{\tau}}(t) \right],\\
	R_{FG}'''(t) &= \tilde{\tau}(1 + t^2)^{-5/2} \left[ -3\tilde{\tau} t \sqrt{1 + t^2} S_{\tilde{\nu}, \tilde{\tau}}(t) + \left( (\tilde{\tau}^2 + 2) t^2 + \tilde{\tau}^2 - 1 \right) C_{\tilde{\nu}, \tilde{\tau}}(t) \right]
\end{align*}
holds for $t \in D_F$.
First, we show the implication $F \leq_{gs}^{t_0} G \Rightarrow \tau_F \geq 2\tau_G$ by contradiction. In order to obtain the asymptotic behaviour of $S_{\tilde{\nu}, \tilde{\tau}}(t)$, we rewrite it as
\begin{align*}
	S_{\tilde{\nu}, \tilde{\tau}}(t) &= \sinh(\tilde{\tau} \cdot \log(t+\sqrt{1+t^2}) - \tilde{\nu})\\
	&= \tfrac{1}{2} \left[ \exp\left(\tilde{\tau} \cdot \log(t+\sqrt{1+t^2}) - \tilde{\nu}\right) - \exp\left(-\tilde{\tau} \cdot \log(t+\sqrt{1+t^2}) + \tilde{\nu}\right) \right]\\
	&= \frac{1}{2} \left[ \frac{(t+\sqrt{1+t^2})^{\tilde{\tau}}}{e^{\tilde{\nu}}} - \frac{e^{\tilde{\nu}}}{(t+\sqrt{1+t^2})^{\tilde{\tau}}} \right].
\end{align*}
Since $\tilde{\tau}>0$, the second summand converges to zero as $t \to \infty$. The first summand is obviously positive and diverges; asymptotically it behaves like $(2t)^{\tilde{\tau}}/e^{\tilde{\nu}}$. Overall, $S_{\tilde{\nu}, \tilde{\tau}}(t) \sim 2^{\tilde{\tau}-1} e^{-\tilde{\nu}} |t|^{\tilde{\tau}}$ for $t \to \infty$. For the asymptotic behaviour as $t \to -\infty$, note that $t+\sqrt{1+t^2}$ behaves for $t \to -\infty$ as
\begin{equation*}
	\sqrt{1+t^2}-t = \frac{(\sqrt{1+t^2}-t)(\sqrt{1+t^2}+t)}{\sqrt{1+t^2}+t} = \frac{1}{\sqrt{1+t^2}+t} \sim (2t)^{-1}
\end{equation*}
does for $t \to \infty$. With similar reasoning as before, we obtain $S_{\tilde{\nu}, \tilde{\tau}}(t) \sim -2^{\tilde{\tau}-1} e^{\tilde{\nu}} |t|^{\tilde{\tau}}$ for $t \to -\infty$. The relationship between the hyperbolic functions now gives $C_{\tilde{\nu}, \tilde{\tau}}(t) \sim S_{\tilde{\nu}, \tilde{\tau}}(t)$ for $t \to \infty$ and $C_{\tilde{\nu}, \tilde{\tau}}(t) \sim -S_{\tilde{\nu}, \tilde{\tau}}(t)$ for $t \to -\infty$. Overall, we infer
\begin{align}
\nonumber
	R_{FG}''(t) &\sim \tilde{\tau} |t|^{-3} \left[ \tilde{\tau} |t| S_{\tilde{\nu}, \tilde{\tau}}(t) - t C_{\tilde{\nu}, \tilde{\tau}}(t) \right] \sim \tilde{\tau}(\tilde{\tau}-1) \frac{S_{\tilde{\nu}, \tilde{\tau}}(t)}{|t|^2}\\
\label{eqn:leq3SAS2ndAblAsymp}
	&\sim \begin{cases}
		\tilde{\tau} (\tilde{\tau}-1) 2^{\tilde{\tau}-1} e^{-\tilde{\nu}} |t|^{\tilde{\tau}-2} &\quad \text{for } t \to \infty,\\
		-\tilde{\tau} (\tilde{\tau}-1) 2^{\tilde{\tau}-1} e^{\tilde{\nu}} |t|^{\tilde{\tau}-2} &\quad \text{for } t \to -\infty,
	\end{cases}
\end{align}
if $\tilde{\tau} \neq 1$. In the case $\tilde{\tau} = 1$, the asymptotically leading summands of $\sqrt{1+t^2} S_{\tilde{\nu}, \tilde{\tau}}(t)$ and $t C_{\tilde{\nu}, \tilde{\tau}}(t)$ cancel out and, therefore, a closer investigation is required. Specifically,
\begin{align*}
	&\phantom{=} \; \tilde{\tau} \sqrt{1+t^2} S_{\tilde{\nu}, \tilde{\tau}}(t) - t C_{\tilde{\nu}, \tilde{\tau}}(t)\\
	&= \frac{\sqrt{1+t^2}}{2} \left[ \frac{(t+\sqrt{1+t^2})^{\tilde{\tau}}}{e^{\tilde{\nu}}} - \frac{e^{\tilde{\nu}}}{(t+\sqrt{1+t^2})^{\tilde{\tau}}} \right] - \frac{t}{2} \left[ \frac{(t+\sqrt{1+t^2})^{\tilde{\tau}}}{e^{\tilde{\nu}}} + \frac{e^{\tilde{\nu}}}{(t+\sqrt{1+t^2})^{\tilde{\tau}}} \right]\\
	&= \frac{1}{2} \left[ \frac{\sqrt{1+t^2}^2 - t^2}{e^{\tilde{\nu}}} - \frac{(t+\sqrt{1+t^2}) e^{\tilde{\nu}}}{t + \sqrt{1+t^2}} \right]\\
	&= \frac{1-e^{2\tilde{\nu}}}{2e^{\tilde{\nu}}}
\end{align*}
yields
\begin{equation}
\label{eqn:leq3SAS2ndAblAsympTau=1}
	R_{FG}''(t) = \frac{1-e^{2\tilde{\nu}}}{2e^{\tilde{\nu}}} (1+t^2)^{-3/2} \sim \frac{1-e^{2\tilde{\nu}}}{2e^{\tilde{\nu}}} |t|^{-3}
\end{equation}
for $\tilde{\tau} = 1$.
Now, assuming $\tilde{\tau} < 2$, it follows that $R_{FG}''(t) \stackrel{|t| \to \infty}{\to} 0$. If $t_0 > 0$, $R_{FG}''(t_u) < t_0$ follows for $t_u$ large enough. However, since $t_0$ lies in the interior of the image of $R_{FG}''$, there also exists a $t_\ell < t_u$ such that $R_{FG}''(t_\ell) > t_0$. This contradicts $F \leq_{gs}^{t_0} G$. If $t_0 < 0$, $R_{FG}''(s_\ell) > t_0$ follows for $s_\ell$ small enough and, by assumption, there also exists an $s_u > s_\ell$ such that $R_{FG}''(s_u) < t_0$, thus also contradicting $F \leq_{gs}^{t_0} G$.\par
We now prove the implication $\tau_F \geq 2\tau_G \Rightarrow F \leq_3 G$ and therefore assume $\tilde{\tau} \geq 2$. $F \leq_3 G$ is equivalent to
\begin{equation}
\label{eqn:SASleq3RevImplCond}
	[(\tilde{\tau}^2 + 2) t^2 + \tilde{\tau}^2 - 1] C_{\tilde{\nu}, \tilde{\tau}}(t) \geq 3 \tilde{\tau} t \sqrt{1+t^2} S_{\tilde{\nu}, \tilde{\tau}}(t)
\end{equation}
holding for all $t \in \R$. Because of $C_{\tilde{\nu}, \tilde{\tau}}(t) \geq 0$ and $(\tilde{\tau}^2+2) t^2 + \tilde{\tau}^2 - 1 \geq 6t^2+3>0$, the left hand side of inequality (\ref{eqn:SASleq3RevImplCond}) is positive for all $t$. Hence, substituting both sides of the inequality with their squares gives a sufficient condition. We obtain
\begin{align*}
	&\phantom{\Leftrightarrow} \hspace{1.45mm} \left[(\tilde{\tau}^2+2) t^2 + (\tilde{\tau}^2-1)\right]^2 C_{\tilde{\nu}, \tilde{\tau}}^2(t) \geq 9 \tilde{\tau}^2 t^2 (1+t^2) (C_{\tilde{\nu}, \tilde{\tau}}^2(t) - 1) \hspace{2.69cm} \forall t \in \R\\
	&\Leftrightarrow \left[ \left( (\tilde{\tau}^2+2)^2 - 9\tilde{\tau}^2 \right) t^4 + \left( 2(\tilde{\tau}^2+2)(\tilde{\tau}^2-1) - 9\tilde{\tau}^2 \right) t^2 + (\tilde{\tau}^2-1)^2 \right] C_{\tilde{\nu}, \tilde{\tau}}^2(t)\\
	&\phantom{\Leftrightarrow} \hspace{9.36cm} + 9 \tilde{\tau}^2 t^2 (1+t^2) \geq 0 \ \forall t \in \R.
\end{align*}
The second summand on the left hand side is obviously non-negative. It is now sufficient to show that all coefficients of the polynomial, with which $C_{\tilde{\nu}, \tilde{\tau}}^2(t)$ is multiplied, are non-negative. For the constant $(\tilde{\tau}^2-1)^2$, this is obvious. The coefficient of $t^2$ is equal to $2\tilde{\tau}^4 - 7\tilde{\tau}^2-4 = (\tilde{\tau}^2-4)(2\tilde{\tau}^2+1)$, which is non-negative since $\tilde{\tau}\geq 2$ was assumed. The same is true for the coefficient of $t^4$, which equals $\tilde{\tau}^4-5\tilde{\tau}^2+4 = (\tilde{\tau}^2-4)(\tilde{\tau}^2-1)$. This concludes the proof of the chain (\ref{eqn:leq3gsSASChain}) of implications.

It remains to prove the equivalent characterization of $F \leq_{gs}^0 G$, so let now $t_0 = 0$. Note that the sign of $R_{FG}''(t)$ corresponds to the sign of $h(t) = \tilde{\tau} \sqrt{1+t^2} S_{\tilde{\nu}, \tilde{\tau}}(t) - t C_{\tilde{\nu}, \tilde{\tau}}(t), t \in \R$.
%First, we assume that $\tilde{\tau} > 1$.
Using (\ref{eqn:leq3SAS2ndAblAsymp}), (\ref{eqn:leq3SAS2ndAblAsympTau=1}) as well as $h(t) = (1+t^2)^{3/2}R_{FG}''(t)/\tilde{\tau} , t \in \R$, we obtain that
\begin{equation}
\label{eqn:leq3SAS2ndAblAsympPartial}
	h(t) \sim \begin{cases}
		(\tilde{\tau}-1) 2^{\tilde{\tau}-1} e^{-\tilde{\nu}} |t|^{\tilde{\tau}+1} \quad &\text{for } t \to \infty,\\
		-(\tilde{\tau}-1) 2^{\tilde{\tau}-1} e^{\tilde{\nu}} |t|^{\tilde{\tau}+1} \quad &\text{for } t \to -\infty,
	\end{cases}
\end{equation}
for $\tilde{\tau} \neq 1$ and
\begin{equation*}
	h(t) = \frac{1 - e^{2\tilde{\nu}}}{2e^{\tilde{\nu}}}, \quad t \in \R,
\end{equation*}
for $\tilde{\tau} = 1$. From the latter, we infer that either $R_{FG}'' \geq 0$ (in the case $\tilde{\nu}<0$) or $R_{FG}'' \leq 0$ (in the case $\tilde{\nu}>0$) holds. (Note that the case $\tilde{\nu}=0$ is excluded due to the assumption $F \neq G$.) While this yields $F \leq_{gs}^0 G$ for $\tilde{\tau} = 1$, the threshold $t_0 = 0$ does not satisfy $t_0 \in \mathrm{int}(R_{FG}''(D_F))$, which is assumed in the result.
Continuing under the assumption $\tilde{\tau} \neq 1$, (\ref{eqn:leq3SAS2ndAblAsympPartial}) yields
\begin{equation}
\label{eqn:leq3SAS2ndAblGWPartial}
	\lim_{t \to \pm \infty} h(t) = \begin{cases}
		\pm \infty \quad &\text{for } \tilde{\tau} > 1,\\
		%\frac{1 - e^{2\tilde{\nu}}}{e^{\tilde{\nu}}} \quad &\text{for } \tilde{\tau} = 1,\\
		\mp \infty \quad &\text{for } \tilde{\tau} < 1.
	\end{cases}
\end{equation}
Considering $S'_{\tilde{\nu}, \tilde{\tau}}(t) = \tilde{\tau} (1+t^2)^{-1/2} C_{\tilde{\nu}, \tilde{\tau}}(t)$ and $C'_{\tilde{\nu}, \tilde{\tau}}(t) = \tilde{\tau} (1+t^2)^{-1/2} S_{\tilde{\nu}, \tilde{\tau}}(t)$, the derivative of $h$ is given by
\begin{align*}
	h'(t) &= \tilde{\tau} t (1+t^2)^{-1/2} S_{\tilde{\nu}, \tilde{\tau}}(t) + \tilde{\tau} (1+t^2)^{1/2} S'_{\tilde{\nu}, \tilde{\tau}}(t) - C_{\tilde{\nu}, \tilde{\tau}}(t) - t C'_{\tilde{\nu}, \tilde{\tau}}(t)\\
	&= \tilde{\tau} t (1+t^2)^{-1/2} S_{\tilde{\nu}, \tilde{\tau}}(t) + \tilde{\tau}^2 C_{\tilde{\nu}, \tilde{\tau}}(t) - C_{\tilde{\nu}, \tilde{\tau}}(t) - \tilde{\tau} t (1+t^2)^{-1/2} S_{\tilde{\nu}, \tilde{\tau}}(t)\\
	&= (\tilde{\tau}^2 - 1) C_{\tilde{\nu}, \tilde{\tau}}(t)
\end{align*}
for $t \in \R$. Because of $C_{\tilde{\nu}, \tilde{\tau}}(t) > 0$, we have $h' > 0$ for $\tilde{\tau}>1$ and $h' < 0$ for $\tilde{\tau}<1$. Combined with (\ref{eqn:leq3SAS2ndAblGWPartial}), it follows that $h$ has exactly one root, at which its sign changes from '$-$' to '$+$' if $\tilde{\tau} > 1$ and from '$+$' to '$-$' if $\tilde{\tau} < 1$. Since the sign of $R_{FG}''$ coincides with the sign of $h$, it follows directly that $F \leq_{gs}^{0} G$ holds for $\tilde{\tau}>1$ and that the same does not hold for $\tilde{\tau}<1$.

\begin{remark}
	It follows from $F \neq G$ that $\mathrm{int}(R_{FG}''(D_F)) \neq \emptyset$, so that the equivalent characterizations of $F \leq_{gs}^{t_0}$ in Theorem 4 are not statements about the empty set. We prove this by contradiction and therefore assume $\mathrm{int}(R_{FG}''(D_F)) = \emptyset$. Since $R_{FG}''$ is continuous, this occurs if and only if $R_{FG}''$ is constant. Defining the function $h(t) = \tilde{\tau} \sqrt{1+t^2} S_{\tilde{\nu}, \tilde{\tau}}(t) - t C_{\tilde{\nu}, \tilde{\tau}}(t), t \in \R,$ as in the proof of Theorem 4, this is equivalent to the existence of a constant $c \in \R$ such that $h(t) = c (1+t^2)^{3/2}, t \in \R$. The case $c=0$ is equivalent to $F=G$ as $h$ is not constant for $\tilde{\tau}\neq 1$ and non-zero for $\tilde{\tau}=1$ and $\tilde{\nu} \neq 0$. In the case $c \neq 0$, we either obtain $\lim_{t \to \pm \infty} h(t) = \infty$ (for $c>0$) or $\lim_{t \to \pm \infty} h(t) = -\infty$ (for $c<0$), which contradicts (\ref{eqn:leq3SAS2ndAblGWPartial}) in combination with the fact that $h$ is constant for $\tilde{\tau}=1$.
\end{remark}

\section{Behaviour of the functions $\boldsymbol{R_{FG}}$ for sinh-arsinh distributions}

\begin{figure}
	\centering{
\includegraphics[scale=0.65]{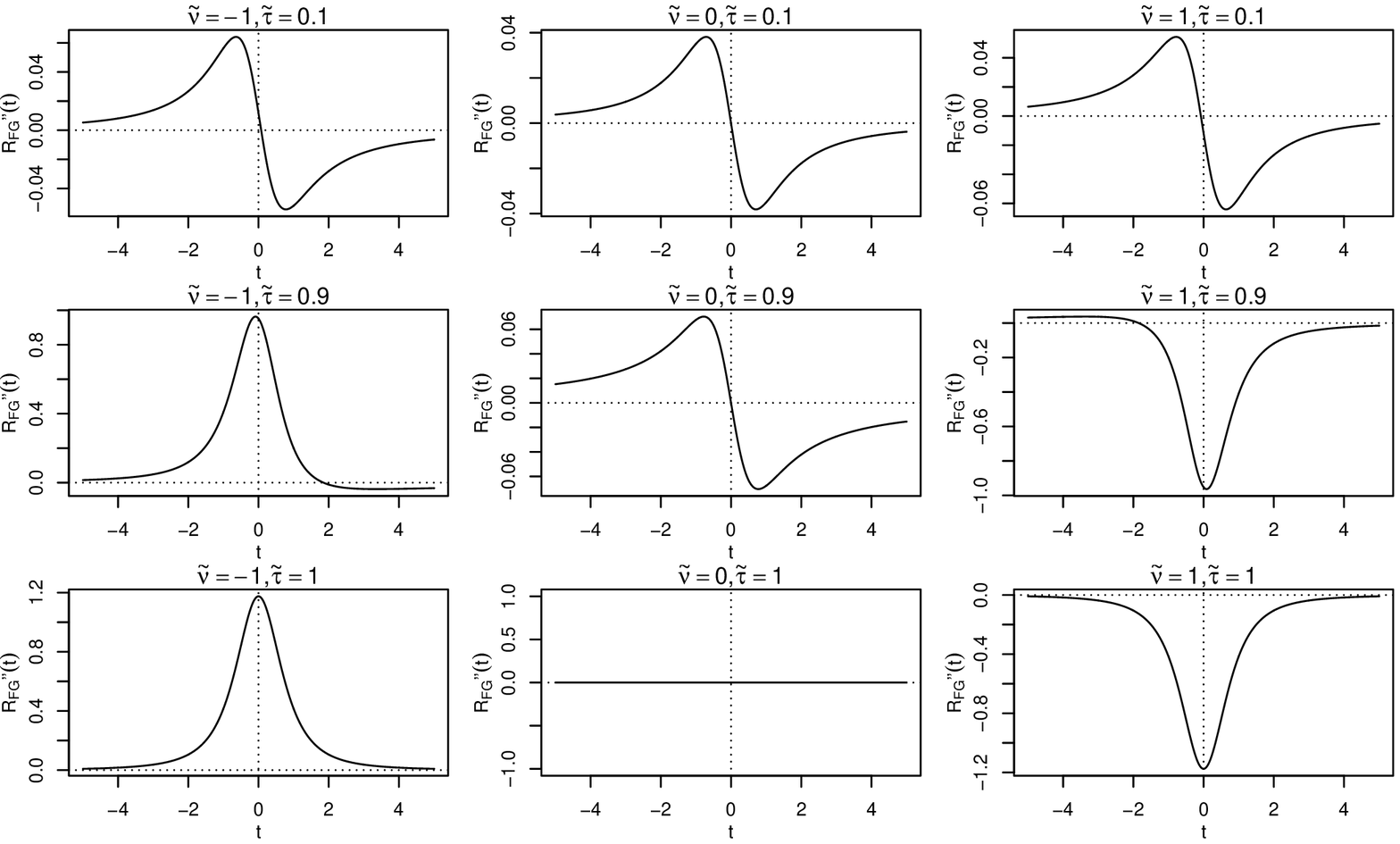}
		\caption{\label{fig:scndAblRIDF_SAS1}Graphs of $R_{FG}''$  with $F$ and $G$ being the cdf's of $X \sim \SAS(\nu_F, \tau_F)$ and $Y \sim \SAS(\nu_G, \tau_G)$, respectively.}
	}
\end{figure}

\begin{figure}
	\centering{
		\includegraphics[scale=0.65]{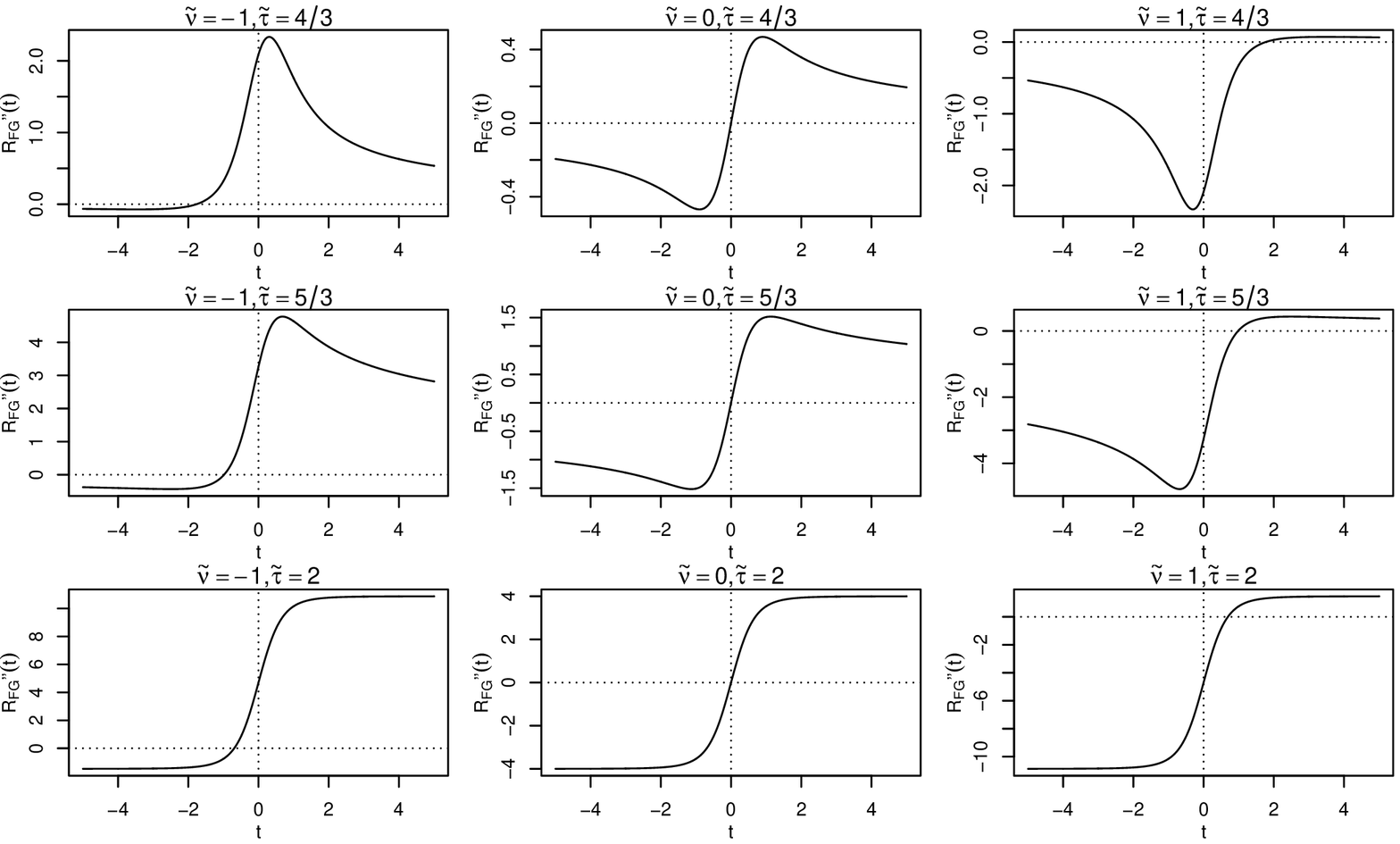}
		\caption{\label{fig:scndAblRIDF_SAS2}Graphs of $R_{FG}''$  with $F$ and $G$ being the cdf's of $X \sim \SAS(\nu_F, \tau_F)$ and $Y \sim \SAS(\nu_G, \tau_G)$, respectively.}
	}
\end{figure}

\begin{figure}
	\centering{
		\includegraphics[scale=0.65]{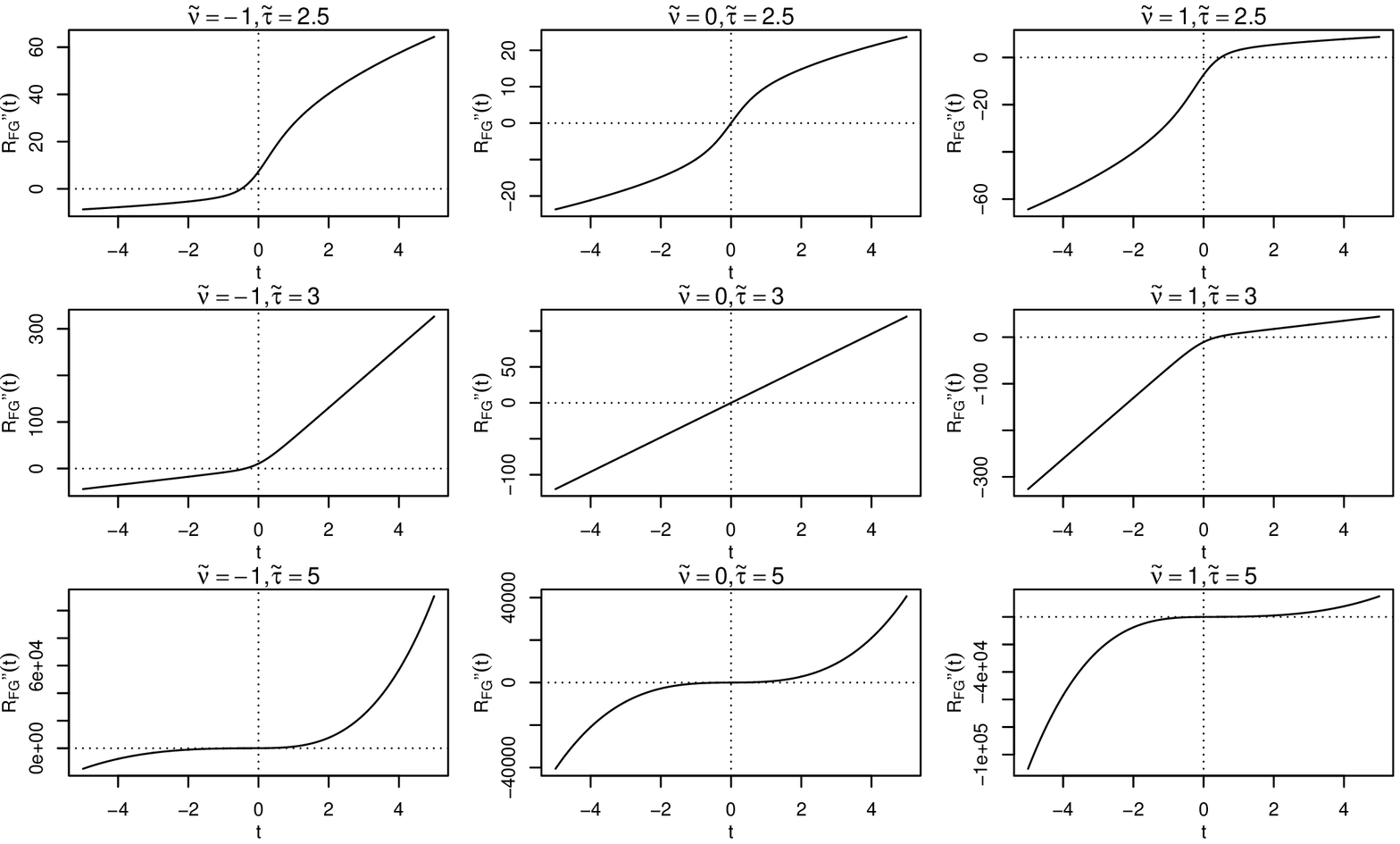}
		\caption{\label{fig:scndAblRIDF_SAS3}Graphs of $R_{FG}''$  with $F$ and $G$ being the cdf's of $X \sim \SAS(\nu_F, \tau_F)$ and $Y \sim \SAS(\nu_G, \tau_G)$, respectively.}
	}
\end{figure}

For a number of choices of $\tilde{\nu}$ and $\tilde{\tau}$, the function $R_{FG}''$ is plotted in Figures \ref{fig:scndAblRIDF_SAS1}, \ref{fig:scndAblRIDF_SAS2} and \ref{fig:scndAblRIDF_SAS3}. Additionally, a number of properties are summarized in Table 1 in the main paper. It is obvious from (\ref{eqn:leq3SAS2ndAblAsymp}) and (\ref{eqn:leq3SAS2ndAblAsympPartial}) that $R_{FG}''$ asymptotically always behaves like a monomial, where the exponent is linearly increasing in $\tilde{\tau}$ (except for the case $\tilde{\tau} = 1$). The exponent reaches the value $0$ for $\tilde{\tau}=2$, which corresponds to the fact that $F \leq_3 G$ is equivalent to $\tilde{\tau}\geq 2$. Furthermore, the function has exactly one root for $\tilde{\tau} \neq 1$ with the direction of the sign change switching for $\tilde{\tau} = 1$. The graph of $R_{FG}''$ is point symmetric around the origin for $\tilde{\nu} = 0$. For $\tilde{\nu} < 0$, the side with the positive values of $R_{FG}''$ is scaled up and the other side is scaled down. Additionally, the sole root of the function shifts to the side with the scaled-down values. The reverse is true for $\tilde{\nu}>0$ with the extent of the rescaling and the shift exponentially depending on the absolute value of $\tilde{\nu}$.\par
In the symmetric case of $\tilde{\nu} = 0$, a number of special cases stand out, which are also singled out in Table 1. First, for $\tilde{\tau}=1$, $R_{FG}'' \equiv 0$ obviously holds since $F=G$ and, therefore, $R_{FG}$ is the identity function (see lower central panel of Figure \ref{fig:scndAblRIDF_SAS1}). Then, for $\tilde{\tau} = 2$, the rather simple form $R_{FG}(t) = 2t \sqrt{t^2 + 1}$ is obtained, yielding the second derivative $R_{FG}''(t) = (4t^3 + 6t)(t^2+1)^{-3/2}$, which converges to $4$ as $t \to \infty$ and to $-4$ as $t \to -\infty$ (see lower central panel of Figure \ref{fig:scndAblRIDF_SAS2}). Finally, for $\tilde{\tau} = 3$, the RIDF is given by $R_{FG}(t) = 4t^3 + 3t$, which leads to the linear second derivative $R_{FG}''(t) = 24t$ (see central panel of Figure \ref{fig:scndAblRIDF_SAS3}).

\end{document}